\documentclass[%
reprint,
amsmath,amssymb,
aps,
]{revtex4-1}

\usepackage{graphicx}
\usepackage{dcolumn}
\usepackage{bm}

\begin{document}

    \preprint{APS/123-QED}




\title{\large\bf Fano effect in Aharonov-Bohm ring
 with topologically superconducting bridge}

\author{V.\,V.\, Val'kov$^{1}$}
\email{vvv@iph.krasn.ru}
\author{M.\,Yu.\,Kagan$^{2,3}$}
\email{kagan@kapitza.ras.ru}
\author{S.V. Aksenov$^{1}$}%
\email{asv86@iph.krasn.ru}
\affiliation{%
    $^1$ Kirensky Institute of Physics, Federal Research Center KSC SB RAS, 660036 Krasnoyarsk, Russia\\%
    $^2$ P.L. Kapitza Institute for Physical Problems RAS, 119334 Moscow, Russia\\
    $^3$ National Research University Higher School of Economics, 101000 Moscow, Russia}

\date{\today}

\begin{abstract}

Taking into account an inner structure of the arms of the Aharonov-Bohm ring (AB ring) we have analyzed the transport features related to the topological phase transition which is induced in a superconducting wire (SC wire) with strong spin-orbit interaction (SOI). The SC wire acts as a bridge connecting the arms. The in-plane magnetic-field dependence of linear-response conductance obtained using the nonequilibrium Green's functions in the tight-binding approximation revealed the Breit-Wigner and Fano resonances (FRs) if the wire is in the nontrivial phase. The effect is explained by the presence of two interacting transport channels in the system. As a result, the FRs are attributed to bound states in continuum (BSCs). The BSC lifetime is determined by both hopping parameters between subsystems and the SC-wire properties. It is established that the FR width and position are extremely sensitive to the type of the lowest-energy excitation in the SC wire, the Majorana or Andreev bound state (MBS or ABS, respectively). Moreover, it is shown that in the specific case of the AB ring, the T-shape geometry, the FR disappears for the transport via the MBS and the conductance is equal to one quantum. It doubles in the local transport regime. On the contrary, in the ABS case the local conductance vanishes.  The influence of the mean-field Coulomb interactions and diagonal disorder in the SC wire on the FR is investigated.

    \begin{description}
        \item[PACS number(s)]
        71.10.Pm, 
        74.45.+c, 
        74.78.Na, 
        85.75.-d 
    \end{description}
\end{abstract}

\maketitle


\section{\label{sec1}Introduction}

In the last decades the development of paradigm concerning the band-structure topological properties exhibits serious achievements \cite{qi-11,bansil-16,armitage-18}. Among all the variety of solid-state systems where the topological phases are found one-dimensional (1D) or quasi-1D  topological superconductors (SCs) attract interest since they provide the basis for novel quantum-electronic bits \cite{elliott-15}. In these structures, by analogy with topological insulators, the bulk SC gap coexists with edge state. It was originally shown in the Kitaev model \cite{kitaev-01} that this state can be defined by two unpaired Majorana fermions (MFs) or Majorana bound state (MBS) localized at the opposite ends of 1D wire. The spatial separation is a key feature allowing to propose MBS-based bits as building blocks of topological quantum computers, in turn, making them stable against local decoherence processes. Additionally, the MFs are non-Abelian anions then the state of such a topological qubit can be changed by means of braiding operations \cite{kitaev-03,alicea-11}.

To induce the MBS in 1D wire the combination of spin-orbit interaction (SOI), magnetic field and SC pairing is necessary \cite{lutchyn-10,oreg-10}. Following this recipe many experiments were done to detect the MBS in a last few years. In particular, tunneling spectroscopy of SC-covered semiconducting wires with strong SOI (in what follows we call this hybrid structure a SC wire) was performed to observe a zero-bias peak (ZBP) in conductance as the MBS fingerprint \cite{mourik-12,deng-12,das-12}. Due to the fact that Majorana particle and antiparticle are the same object the resonant Andreev reflection in the lead-wire contact has to occur guaranteeing the height of ZBP equal to $2G_{0}$ (where $G_{0}=e^2/h$ - conductance quantum) \cite{law-09,flensberg-10} and persistent under the chemical potential, magnetic field and tunnel barrier variations \cite{wu-12,dassarma-12,rainis-13}.

Technological imperfections remained a stable quantization at $2G_{0}$ unattainable \cite{mourik-12,deng-12,das-12,albrecht-16,deng-16,nichele-17} stimulating a wide discussion in the field about both the reasons of low ZBP and other issues resulting in the ZBP and mimicking the MBSs. The saturation of $2G_{0}$ ZBP was prevented by proximity-induced soft rather than hard SC gap. There were different mechanisms suggested to be responsible for the SC phase breaking: disorder in the wire, temperature effect and quasiparticle broadening as a result of lead-wire interface roughness, tunnel barrier fluctuations, electron-electron scattering in the wire as well as interactions with electrons and phonons in the leads \cite{takei-13,liu-17a}. It was shown that disorder in multiband quasi-1D wire can generate not quantized ZBP due to the appearance of either edge \cite{liu-12} or bulk \cite{bagrets-12} quasiparticles. Similar conductance peculiarity emerges due to the Kondo resonance and coalescing Andreev bound states (ABSs) \cite{lee-12,kells-12b,stanescu-13}. According to the geometry of some tunneling spectroscopy experiments these states are able to arise in confined quantum-dot(QD)-like area between normal lead and SC wire \cite{kells-12b,lee-14,deng-16,liu-17b,liu-18}. In addition, such QDs can occur in the wire bulk at the expense of inhomogeneities. However, recent advances in epitaxial growth of hybrid SC wires allowed to achieve the hard induced gap \cite{chang-15} and, finally, reach ballistic transport regime \cite{zhang-17} causing persistent conductance quantization at $2G_{0}$ in the topologically nontrivial phase \cite{zhang-18,franz-18}.

Starting from the seminal study of Aharonov and Bohm \cite{aharonov-59} the influence of distant electromagnetic potential on quantum particle propagation has been utilized in numerous works to research coherent transport in mesoscopic structures (see e.g. \cite{yacoby-94,kobayashi-03,cabosart-14}). The Aharonov-Bohm (AB) effect in different geometries is actively employed to investigate the MBS features as well. The classic one implies the SC wire inserted into one of the AB-ring arms. In particular, the topological phase transition is signalized by a disorder-independent period doubling of the AB conductance oscillations since the MBS allows to transfer individual quasiparticle (not a Cooper pair) \cite{akhmerov-11}. Similarly, it causes the fractional Josephson effect ($4\pi$-periodic) \cite{kitaev-01,snelder-13} and the appearance of $h/e$ harmonic of persistent current finite at zero AB flux with a sign that is determined by the fermion parity of Coulomb-blockaded SC wire \cite{jacquod-13}. The floating SC wire was also used to point out the differences in the AB interference pictures mediated by MBSs and ABSs \cite{hell-18}. The same problem was considered in \cite{tripathi-16} where the conductance and noise were calculated for the AB ring including only one MBS or ABS.

Non-standard geometries include MF-QD-MF connection with the loop-shaped SC wire. Such a scheme is proposed to realize the tunnel-braid topological qubit. Then the non-Abelian rotations within its degenerate ground-state manifold can be done by single-electron tunneling in and out of the Coulomb-blockaded QD. The magnetic flux penetrating the loop center tunes the MF-dot-MF system to the desired degenerate energy point \cite{flensberg-11,liu-11}. In order to distinguish the topological $h/e$-periodicity from the same AB effect provided by the normal ring, the resembling system with the Coulomb-blockaded SC wire to control the ground-state parity, by analogy with \cite{fu-10}, was studied in \cite{chiu-18}. In \cite{sau-15} the middle QD was changed to 1D lead (called a loop in the article) described by the tight-binding model with no Zeeman and SOI terms and coupled to the normal contact in the central part. It was shown that the nontrivial phase of the SC wire results in $4\pi$ conductance oscillations (if the flux quantum is defined as $\phi_{0}=h/2e$) due to the MBS-assisted nonlocal tunneling process while no AB oscillations were observed in the trivial phase.

One of the features of the AB-ring coherent transport is the Fano effect \cite{fano-61} that is a manifestation of the destructive interference. Several schemes were proposed to detect the MBSs involving this physics. In \cite{dessotti-14} the modification of the Fano resonance (FR) in conductance was studied for the AB ring consisting of two QDs and the Kitaev wire side-coupled to one of them. Shang et al placed the wire between the QDs \cite{shang-14}. The SC wire was also considered as one of the contacts \cite{ueda-14,jiang-15}. The contributions of local and crossed Andreev reflection processes resulting in the FR were additionally analyzed for the different the AB-ring geometries \cite{zhang-18b}.

\begin{figure}[tb]
\includegraphics[width=0.5\textwidth]{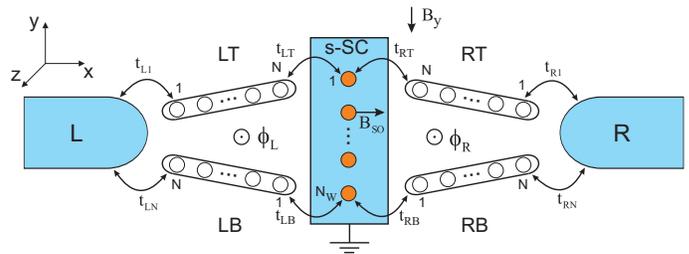}
\caption{\label{1} The superconducting wire embedded in the Aharonov-Bohm ring that is formed by the left top and bottom (LT and LB) as well as right top and bottom (RT and RB) leads. The device is between metallic contacts.}
\end{figure}
Usually, both reference arm of the AB ring and the leads which couple the SC wire (or another structure, e.g. QD) with the contact(s) are skipped or treated phenomenologically. In other words, the normal contact(s) is directly connected with the SC wire or with each other. At the same time, the possibility to fabricate quantum networks based on hybrid superconductor-semiconductor nanowires, including AB rings, was recently demonstrated \cite{gazibegovic-17}. Taking it into account, here we consider lengthy leads in both AB arms, top and bottom, bridged by the SC wire as it is displayed in Fig.\ref{1}. The ring is subjected to an in-plane magnetic field, $B_{y}$. It induces the topological phase transition in the wire and allows to obtain non-topological low-energy excitations in the leads. As a result, the interaction between the transport channels of the wire and leads gives rise to the FRs. Using the nonequilibrium Green's function technique the influence of different factors, such as the AB phase, SC pairing etc., on the FRs is discussed. It is interesting to note that the analogous geometry was investigated in \cite{shang-14}. However, the authors did not treat the leads and SC wire microscopically. The tight-binding treatment of such a scheme was used by Rainis et al \cite{rainis-14} to study the transport signatures of fractional fermions in the wire without focusing on its interaction with the leads' low-energy modes.  

This article is organized as follows: in Sec.\ref{sec2} we present the theoretical model of the system sketched in Fig.\ref{1}. In Sec.\ref{sec3} the theory of quantum transport based on the nonequilibrium Green's function method and tight-binding approximation is discussed. In Sec.\ref{sec4} the results of numerical calculations are analyzed. It includes a general discussion of the observed features of conductance (Sec.\ref{secA}), a comparison of the FR properties for the transport via the MBS and ABS (Sec.\ref{secB}) and the effects of mean-field Coulomb repulsion and diagonal disorder (Sec.\ref{secC}). We conclude in Sec.\ref{sec5} with a summary.

\section{\label{sec2}The model Hamiltonian}

Here we study the wire with the strong Rashba SOI deposited on the surface of s-wave SC. The wire is located between four leads forming the AB-type device (see fig.\ref{1}) with two halfrings. The Hamiltonian of the system depicted in Fig.\ref{1} has the following form:
\begin{equation} \label{H}
\hat{H}=\hat{H}_L+\hat{H}_R+\hat{H}_l+\hat{H}_W+\hat{H}_{Wl}+\hat{H}_{Cl},
\end{equation}
First and second terms in \eqref{H} characterize the left and right paramagnetic metal contacts,
\begin{equation} \label{HLR}
\hat{H}_{i} =\sum\limits_{k\sigma}\left(\xi_{k}\mp\frac{eV}{2}\right)c^+_{ik\sigma}c_{ik\sigma},
~i=L,R,
\end{equation}
where $c_{ik\sigma}$ - an annihilation operator of the electron with momentum $k$, spin $\sigma$ and energy $\xi_{k}=\varepsilon_{k}-\mu$ in the $i$th contact; $\mu$ - a chemical potential of the system; $eV$ - an energy of source-drain electric field. Hereinafter we suppose that a magnetic field, $\mathbf{B}=\left(0,-B_{y},B_{z}\right)$, acts in the device area (leads+SC wire).
The third term, $\hat{H}_{l}=\hat{H}_{LT}+\hat{H}_{LB}+\hat{H}_{RT}+\hat{H}_{RB}$, describes the left-top and -bottom (LT and LB, respectively) as well as right-top and -bottom (RT and RB, respectively) leads which, firstly, connect in parallel the SC wire with the contacts and, secondly, present the arms of AB ring. All the contributions in $\hat{H}_{l}$ are supposed to be identical, e.g.
\begin{eqnarray} \label{Hleads}
&&\hat{H}_{LT} =\sum\limits_{\sigma;j=1}^{N}\xi_{\sigma}d^+_{LTj\sigma}d_{LTj\sigma}-\\
&&~~~~~~~~~~~~~~~~~~~~~~~-\frac{t}{2}\sum\limits_{\sigma;j=1}^{N-1}
\left[d^+_{LTj\sigma}d_{LTj+1,\sigma}+h.c.\right],\nonumber
\end{eqnarray}
where $d_{LTj\sigma}$ - an annihilation operator of the electron on $j$th site of the LT lead with spin $\sigma$ and spin-dependent energy $\xi_{\sigma}=t+\sigma V_{y}-\mu$; $V_{y}=-\frac{1}{2}\mu_B g_e B_{y}$ - $y$-component of the Zeeman energy; $t$ - a nearest neighbor hopping parameter; $g_e$ - an electron $g$-factor. The Hamiltonian of the SC wire can be written as
\begin{eqnarray} \label{HW}
&&\hat{H}_{W} =\sum\limits_{j=1}^{N_{W}}\left[\sum\limits_{\sigma}\xi_{\sigma}a^+_{j\sigma}a_{j\sigma}
 + Un_{j\uparrow}n_{j\downarrow} + \left(\Delta a_{j\uparrow}a_{j\downarrow}+ h.c.\right)\right]\nonumber\\
&&~~~-\sum\limits_{\sigma;j=1}^{N_{W}-1}\left[\frac{1}{2}\left(ta^+_{j\sigma}a_{j+1,\sigma}
+\alpha\sigma a^+_{j\sigma}a_{j+1,\overline{\sigma}}+h.c.\right)+\right.\\
&&~~~~~~~~~~~~~~~~~~~~~~~~~~~~~~~~~~~~~~~~~~~
\left.+V\sum\limits_{\sigma'}n_{j\sigma}n_{j+1,\sigma'}\right],\nonumber
\end{eqnarray}
where $a_{j\sigma}$ - an annihilation operator of the electron on $j$th site of the wire with spin $\sigma$ and energy $\xi_{\sigma}$; $\Delta$ - a SC pairing potential; $\alpha=\alpha_{R}/a$ is a spin-orbit coupling strength obtained via the wire's Rashba parameter, $\alpha_{R}$, and a lattice constant, $a$; $U,~V$ - an intensity of the intra- and intersite Coulomb interactions, respectively; $n_{j\sigma}=a^{+}_{j\sigma}a_{j\sigma}$ - an operator of particle number in the SC wire. The term $\hat{H}_{Wl}$ takes into account the coupling between the SC wire and leads,
\begin{eqnarray} \label{HWl}
&&\hat{H}_{Wl} =\sum\limits_{\sigma}\left[\left(t_{LT}d_{LTN\sigma}^{+}+t_{RT}d_{RTN\sigma}^{+}\right)a_{1\sigma}
\right. + \\
&&~~~~~~~~~~~~~~~+\left.\left(t_{LB}d_{LB1\sigma}^{+}+t_{RB}d_{RB1\sigma}^{+}\right)a_{N_{W}\sigma}\right]+ h.c.,\nonumber
\end{eqnarray}
where $t_{LT(LB)}$, $t_{RT(RB)}$ - overlap integrals between the edges of LT(LB), RT(RB) leads and SC wire, respectively.

The tunnel Hamiltonian $\hat{H}_{Cl}$ is responsible for the coupling between the contacts and leads, which are simultaneously the arms of AB ring,
\begin{eqnarray} \label{HCl}
&&\hat{H}_{Cl} =\sum \limits_{k\sigma}\left[c_{Lk\sigma}^{+}\left(t_{L1}e^{-i\frac{\Phi_L}{2}}d_{LT1\sigma}+
t_{LN}e^{i\frac{\Phi_L}{2}}d_{LBN\sigma}\right)\right. + \nonumber\\
&&+\left. c_{Rk\sigma}^{+}\left(t_{R1}e^{i\frac{\Phi_R}{2}}d_{RT1\sigma} +
t_{RN}e^{-i\frac{\Phi_R}{2}}d_{RBN\sigma}\right)\right]+ h.c.,
\end{eqnarray}
where $t_{L1(N)}$, $t_{R1(N)}$ - overlap integrals between the edges of LT(LB), RT(RB) leads and corresponding contacts; $\Phi_{L\left(R\right)}=2\pi\frac{\phi_{L\left(R\right)}}{\phi_0}$; $\phi_{L\left(R\right)}=B_{z}S_{L\left(R\right)}$ - a magnetic flux threading the left (right) halfring with an area $S_{L\left(R\right)}$; $\phi_0=h/e$ - the flux quantum. Note that the Hamiltonian \eqref{HCl} represents the effective interaction operator which is used to develop quantum-transport theory in the system.

\section{\label{sec3}General quantum-transport theory in the AB ring}

Transport properties of the AB ring is calculated numerically using the nonequilibrium Green's functions \cite{keldysh-65,datta-95,datta-05}. These functions have to be defined in the spin$\otimes$Nambu space to take into account both the SC pairing and spin-flip processes in the wire subsystem \cite{wu-12,wu-14}.

In order to describe how the Green's function technique is combined with the tight-binding approximation in case of the AB ring we introduce the set of basis field operators for $i$th subsystem (the leads, SC wire or contacts) each characterized by its own set of quantum numbers $m$, $\hat{\psi}_{im}=\left(f_{im\uparrow}~f_{im\downarrow}^{+}~ f_{im\downarrow}~f_{im\uparrow}^{+}\right)^T$, where $f_{im\sigma}=\left\{d_{ij\sigma},~a_{j\sigma},~c_{ik\sigma}\right\}$ - an annihilation operator of the electron with quantum number $m$ and spin $\sigma$ acting in the $i$th Hilbert subspace. Specifically, the total field operators of the multi-site leads and SC wire are
\begin{eqnarray} \label{Nambu}
&&\hat{\psi}_{i}=\left(\hat{\psi}_{i1}~...~\hat{\psi}_{iN}\right)^T,~i=LT,LB,RT,RB\nonumber\\
&&\hat{\psi}_{W}=\left(\hat{\psi}_{W1}~...~\hat{\psi}_{WN_{W}}\right)^T.\nonumber
\end{eqnarray}
As a result, the summands in $\hat{H}$ become
\begin{eqnarray} \label{HNambu}
&&\hat{H}_W = \frac{1}{2}\hat{\psi}_{W}^{+}\hat{h}_{W}\hat{\psi}_{W}+const,~
\hat{H}_l = \frac{1}{2}\sum\limits_{i}\hat{\psi}_{i}^{+}\hat{h}_{i}\hat{\psi}_{i},\\
&&\hat{H}_{Wl} =\frac{1}{2}\sum\limits_{k}\left[\hat{\psi}_{W1}^{+}\left(\hat{T}_{LTN}\hat{\psi}_{LTN}+
\hat{T}_{RTN}\hat{\psi}_{RTN}\right)+\right.\nonumber\\
&&~~~~~~~~~~~~~
\left.+
\hat{\psi}_{WN_{W}}^{+}\left(\hat{T}_{LB1}\hat{\psi}_{LB1}+\hat{T}_{RB1}\hat{\psi}_{RB1}\right)+h.c.\right],\nonumber\\
&&\hat{H}_{Cl} =\frac{1}{2}\sum\limits_{k}\left[\hat{\psi}_{Lk}^{+}\left(\hat{T}_{LT1}\hat{\psi}_{LT1}+
\hat{T}_{LBN}\hat{\psi}_{LBN}\right)+\right.\nonumber\\
&&~~~~~~~~~~~~~
\left.+\hat{\psi}_{Rk}^{+}\left(\hat{T}_{RT1}\hat{\psi}_{RT1}+
\hat{T}_{RBN}\hat{\psi}_{RBN}\right)+h.c.\right],\nonumber
\end{eqnarray}
where $\hat{h}_{W}$ - the $4N_{W} \times 4N_{W}$ matrix with the following nonzero blocks without the Coulomb interactions
\begin{eqnarray} \label{HWn}
&&H_{jj}=
\left(\begin{array}{cccc}
\xi_{\uparrow} & \Delta & 0 & 0 \\
\Delta & -\xi_{\downarrow} & 0 & 0 \\
0 & 0 & \xi_{\downarrow} & -\Delta \\
0 & 0 & -\Delta & -\xi_{\uparrow}
\end{array}\right),~\\
&&H_{j,j+1}=H_{j,j-1}=\frac{1}{2}
\left(\begin{array}{cccc}
-t & 0 & -\alpha & 0 \\
0 & t & 0 & -\alpha \\
\alpha & 0 & -t & 0 \\
0 & \alpha & 0 & t
\end{array}\right).\nonumber
\end{eqnarray}
Notice that the corrections to these blocks in case of $U,~V\neq0$ will be discussed in Sec.\ref{secB}.  The blocks of leads' matrix $\hat{h}_{j}$ can be obtained from \eqref{HWn} by putting $\Delta=\alpha=0$.

The tunnel matrices responsible for the contact-lead couplings, $\hat{T}_{LT1,LBN}$ and $\hat{T}_{RT1,RBN}$, are received by performing a gauge transformation \cite{rogovin-74,zeng-03},
\begin{eqnarray} \label{TLR}
&&\hat{T}_{LT1}=t_{L}\hat{\sigma}\hat{V}\hat{\Phi}_{L},~\hat{T}_{LBN}=t_{L}\hat{\sigma}\hat{V}\hat{\Phi}_{L}^{+},\\
&&\hat{T}_{RT1}=t_{R}\hat{\sigma}\hat{V}^{+}\hat{\Phi}_{R},~\hat{T}_{RBN}=t_{L}\hat{\sigma}\hat{V}^{+}\hat{\Phi}_{R}^{+},\nonumber\\
&&\hat{\sigma}=diag\left(1,-1,1,-1\right),\nonumber\\
&&\hat{V}=diag\left(e^{-i\frac{eV}{2}t},e^{i\frac{eV}{2}t},e^{-i\frac{eV}{2}t},e^{i\frac{eV}{2}t}\right),~\nonumber\\
&&\hat{\Phi}_{L\left(R\right)}=diag\left(e^{\mp i\frac{\Phi_{L\left(R\right)}}{2}},e^{\pm i\frac{\Phi_{L\left(R\right)}}{2}},e^{\mp i\frac{\Phi_{L\left(R\right)}}{2}},e^{\pm i\frac{\Phi_{L\left(R\right)}}{2}}\right),~\nonumber
\end{eqnarray}
The tunnel matrices characterizing the wire-lead couplings are $\hat{T}_{LTN\left(LB1\right)}=t_{LT\left(LB\right)}\hat{\sigma}$ and $\hat{T}_{RTN\left(RB1\right)}=t_{RT\left(RB\right)}\hat{\sigma}$.

Let us introduce the matrix nonequilibrium Green's functions in terms of the above-described field operators as
\begin{equation} \label{GF}
\hat{G}_{in,jm}\left(\tau,\tau'\right)=-i\left\langle T_{C}\hat{\psi}_{in}\left(\tau\right)\otimes
\hat{\psi}_{jm}^{+}\left(\tau'\right)\right\rangle,~
\end{equation}
where $T_{C}$ is a Keldysh-contour ordering operator. Next, the current in the left lead is given by $I_{L}=e\left\langle\dot{N}_{L}\right\rangle$ ($N_{L}=\sum_{k\sigma}c_{Lk\sigma}^{+}c_{Lk\sigma}$ is a particle operator in the left lead). And, after some manipulations, we finally find
\begin{eqnarray} \label{IL1}
&&I_{L}=
e\int\limits_{-\infty}^{+\infty}\frac{d\omega}{\pi}
Tr\Biggl[\hat{\sigma}Re\Biggl\{\sum_{\alpha,\beta}\left(
\hat{\Sigma}_{L\alpha\beta}^{r}\hat{G}_{\beta\alpha}^{<} \right.\Biggr.\Biggr.\\
&&\left.\Biggl.\Biggl.~~~~~~~~~~~~~~~~~~~~~~~~~~~~~~~~~~
+\hat{\Sigma}_{L\alpha\beta}^{<}
\hat{G}_{\beta\alpha}^{a}\right) \Biggr\}\Biggr],\nonumber
\end{eqnarray}
where $\hat{\Sigma}_{L\alpha\beta}^{r}$ - the Fourier transform of $\alpha\beta$th block of the matrix self-energy function describing the influence of the left contact on the system, $\hat{\Sigma}_{L\alpha\beta}\left(t-t'\right)=\hat{T}_{\alpha}^{+}\left(t\right)\hat{g}_{Lk}\left(t-t'\right)\hat{T}_{\beta}\left(t'\right)$; $\hat{g}_{Lk}\left(t-t'\right)$ - the free-particle Green's function of the left contact; $\hat{G}_{\beta\alpha}^{<,a}$ - the Fourier transforms of $\beta\alpha$th block of the lesser and advanced Green's functions of the AB ring. The summation in \eqref{IL1} is over the $1$st and $N$th sites of the LT- and LB leads, respectively, i.e. $\alpha,\beta=LT1, LBN$. It means that the current is defined by the contributions of separate top and bottom AB arms ($\alpha=\beta$) and interference terms ($\alpha\neq\beta$). By this fact, the left total self-energy functions are $\hat{\Sigma}_{L}=\sum_{\alpha,\beta}\hat{\Sigma}_{L\alpha\beta}$. Similarly, the summation over the $1$st and $N$th sites of the RT- and RB leads gives the right total self-energy functions, $\hat{\Sigma}_{R}=\sum_{\alpha,\beta}\hat{\Sigma}_{R\alpha\beta}$. Finally, $\hat{\Sigma}=\hat{\Sigma}_{L}+\hat{\Sigma}_{R}$. The retarded, advanced and lesser self-energies are
\begin{eqnarray}
&&\hat{\Sigma}_{i\alpha\beta}^{r\left(a\right)}=\mp\frac{i}{2}\Gamma_{i\alpha\beta},~
\hat{\Sigma}_{i\alpha\beta}^{<}=\left(\hat{\Sigma}_{i\alpha\beta}^{a}-\hat{\Sigma}_{i\alpha\beta}^{r}\right)
\hat{F}_{i},~i=L,R,\nonumber\\
&&\hat{F}_{L\left(R\right)}=diag\biggl(n\left(\omega \pm eV/2\right),~n\left(\omega \mp eV/2\right),\biggr.\\
&&\biggl.~~~~~~~~~~~~~~~~~~~~~~~~~~~~~~~~~~n\left(\omega \pm eV/2\right),~n\left(\omega \mp eV/2\right)\biggr),\nonumber
\end{eqnarray}
where $\Gamma_{i\alpha\beta}=2\pi t_{i\alpha}^2\rho_{i}$ if $\alpha=\beta$ and $\Gamma_{i\alpha\beta}=2\pi t_{i\alpha}t_{i\beta}\rho_{i}$ if $\alpha\neq\beta$ - the coupling strengths between the $i$th leads and the same contact; $\rho_{i}$ - the DOS of the $i$th contact. In the calculations below the contacts are treated in the wide-band limit. It means that $\Gamma_{i\alpha\beta}=const$.

The retarded and lesser Green's functions can be obtained from the Dyson and Keldysh equations, respectively,
\begin{eqnarray}
&&\hat{G}^{r}=\left(\omega-\hat{H}-\hat{\Sigma}^{r}\right)^{-1},~
\hat{G}^{a}=\left(\hat{G}^{r}\right)^{+},~\label{Gr}\\
&&\hat{G}^{<}=\hat{G}^{r}\hat{\Sigma}^{<}\hat{G}^{a}.~\label{G+-}
\end{eqnarray}

\section{\label{sec4}Results and discussion}

In numerical calculations we suppose that $S_L=S_R=S_{\triangle}=\frac{L_{W}}{2}\sqrt{L^2-\frac{L_{W}^2}{4}}$, where $L_{W}=a\left(N_{W}-1\right)$, $L=a\left(N-1\right)$, $N_{W}=30$, $N=20$. It is important to note that we use small $N_{W}$ and $N$ which considerably reduces tight-binding transport calculations. Consequently, in order to get the MBSs in such a short SC wire it is necessary to consider unphysically large lattice constant $a=50$ nm leading to small hopping parameter, $t=\frac{\hbar^2}{m^{*}a^2}$, where $m^{*}=0.015m_{0}$ is an experimentally accessed effective mass in semiconducting nanowires \cite{mourik-12}. The other parameters are $k_{B}T=0$, $\mu=0$, $\Delta=250~\mu eV$, $\alpha_R=0.2~eV\cdot{\AA}$, $g=50$, $B_{y}\sim 0.1-1~T$, $B_{z}\sim 0.001-0.01~T$. In the forthcoming paragraphs, all the energy parameters are measured in units of $t$. The transport parameters satisfy an inequality, $\Gamma_{i},~t_{j} \ll t$, where $i=L1,LN,R1,RN$, $j=LT,LB,RT,RB$, that corresponds to weak (tunnel) coupling. In addition, the linear response regime is studied in the article.
\begin{figure}[htbp]
    \includegraphics[width=0.45\textwidth]{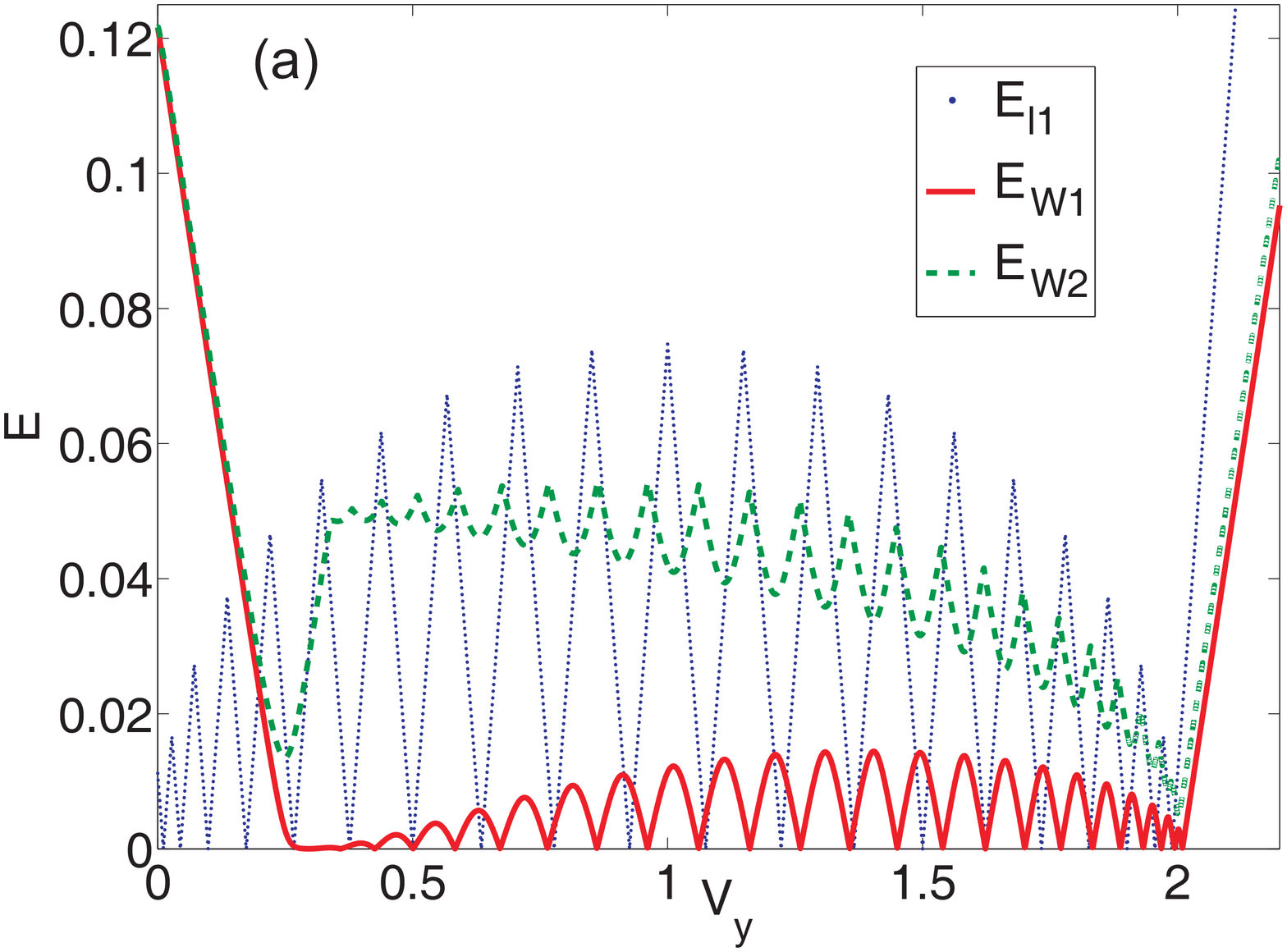}
    \includegraphics[width=0.45\textwidth]{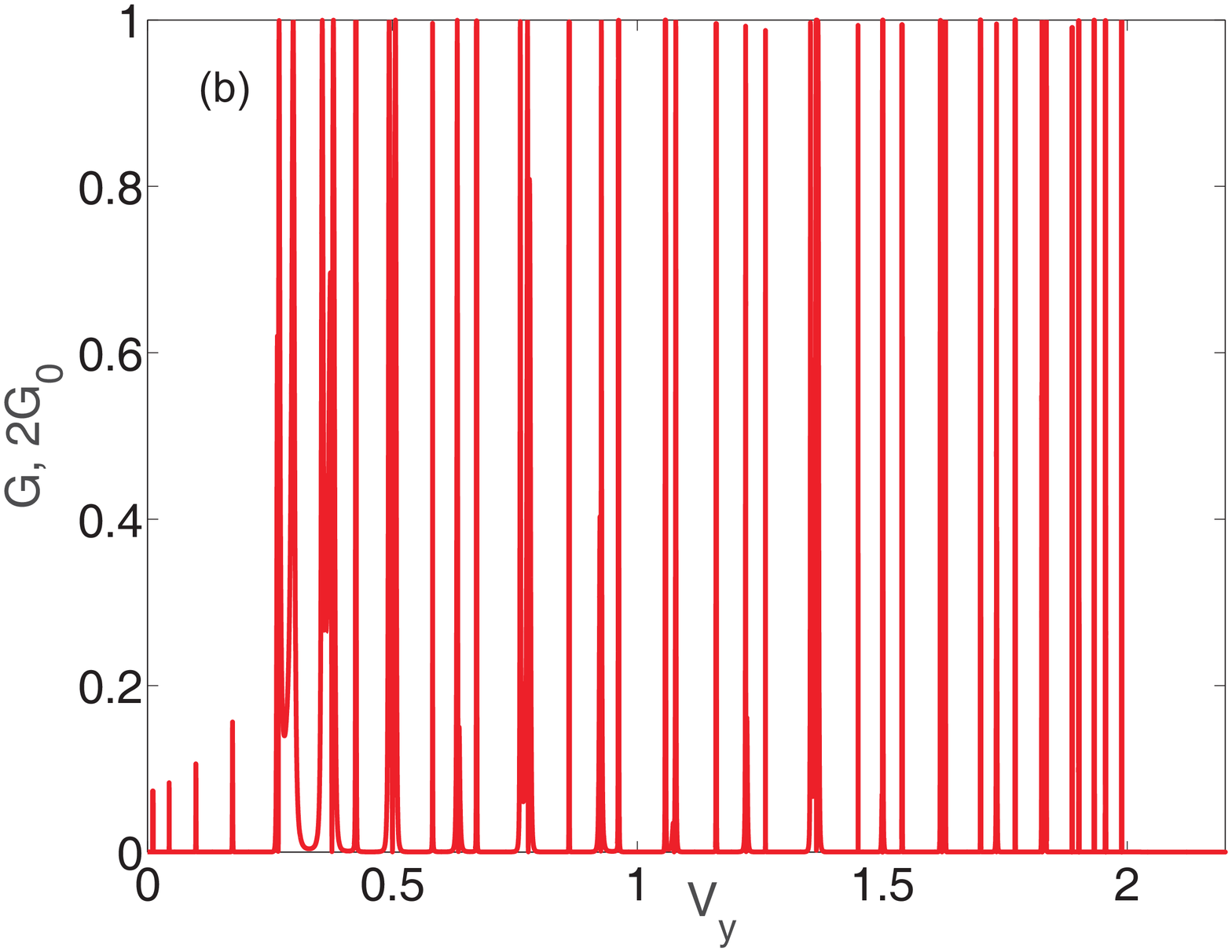}
    \caption{\label{2} The magnetic-field dependence of the lowest energies of the single-electron excitations of the lead and SC wire (a) and the conductance of the AB ring (b). Parameters: $U=V=0$, $\alpha=0.195$, $\Delta=0.243$, $\Phi=0$.}
\end{figure}

\subsection{\label{secA}Breit-Wigner and Fano resonances}

We start with the situation when the Coulomb interactions in the SC wire are neglected ($U=V=0$) and consider the symmetric AB ring, i.e. $\Gamma_{L1}=\Gamma_{LN}=\Gamma_{R1}=\Gamma_{RN}=\Gamma=0.01$, $t_{LT}=t_{LB}=t_{RT}=t_{RB}=t_{1}=0.1$. Initially the AB phases in the halfrings are chosen to be the same, $\Phi_{L}=\Phi_{R}=\Phi$. Hereinafter, we mostly concentrate on an in-plane Zeeman-energy dependence of the conductance since from experimental point of view it is easier to control the magnetic field in comparison with the gate voltage, SOI or SC pairing parameters. The transport in the system is defined by the one-particle excitations in the vicinity of chemical potential. The corresponding energies of the separate lead (blue dotted curve) and SC wire (red solid and green dashed curves) as functions of the in-plane Zeeman energy are shown in Fig.\ref{2}a. The lowest energy of the lead, $E_{l1}$, periodically becomes equal to zero if $V_{y}\lesssim2$. Simultaneously, the topologically nontrivial phase develops in the SC wire if $\mu^2+\Delta^2<V_{y}^2<\left(2t - \mu\right)^2+\Delta^2$ \cite{lutchyn-10,oreg-10}. In this regime the lowest energy of wire excitation, $E_{W1}$, is split from the higher one, $E_{W2}$, and oscillates as it is clearly seen in Fig.\ref{2}a. As a result, the zeros of $E_{W1}$ indicating the appearance of the Majorana mode in the SC wire coexist with the zeros of $E_{l1}$.

The in-plane magnetic-field dependence of conductance is shown in Fig.\ref{2}b. At low magnetic fields, $V_{y}\lesssim0.25$, the peaks of $G\left(V_{y}=E_{l1}\right)$ are significantly suppressed since there is no conducting state in the wire, $E_{W1}\gg0$ (see Fig.\ref{2}b). Additionally, $G\left(V_{y}=E_{l1}\right)\rightarrow1$ if e.g. $\Phi\neq0$ that points out the significant impact of the real-space interference processes. When the SC wire undergoes the topological phase transition the set of conductance resonances appears at $0.25 \lesssim V_{y} \lesssim 2$. At high fields, $V_{y}\gtrsim2$, the linear-response conductance is close to zero as $E_{l1},~E_{W1}\gg0$.

Let us thoroughly consider the features of conductance related to the SC-wire topological phase. All the resonances of $G$ at the fields $0.25 \lesssim V_{y} \lesssim 2$ can be divided into two groups. The first is Breit-Wigner resonances (BWRs) and the second is FRs. To understand deeper the nature of these resonant features in the conductance we firstly analyze a simplified situation when each lead has just one site and the wire consists of two sites. Thus, there is a sextuple-QD structure including two arms coupled in-parallel with the contacts. In Fig.\ref{1} each arm consists of one orange and two white circles. Since the magnetic field results in primarily spin-polarized transport in the original system a spinless regime in the sextuple-QD structure is considered. Denoting the hopping between two central QDs as $t_0$ we receive the following energy spectrum of the sextuple-QD structure:
\begin{eqnarray}\label{spec6QD}
&&E_{1,2}=\xi_{\downarrow},~E_{3,4}=\xi_{\downarrow}+\frac{1}{4}\sqrt{t_0^2+8t_{1}^2}\pm \frac{t_0}{4},\nonumber\\
&&E_{5,6}=\xi_{\downarrow}-\frac{1}{4}\sqrt{t_0^2+8t_{1}^2}\pm \frac{t_0}{4}.
\end{eqnarray}

\begin{figure}[htbp]
	\includegraphics[width=0.45\textwidth]{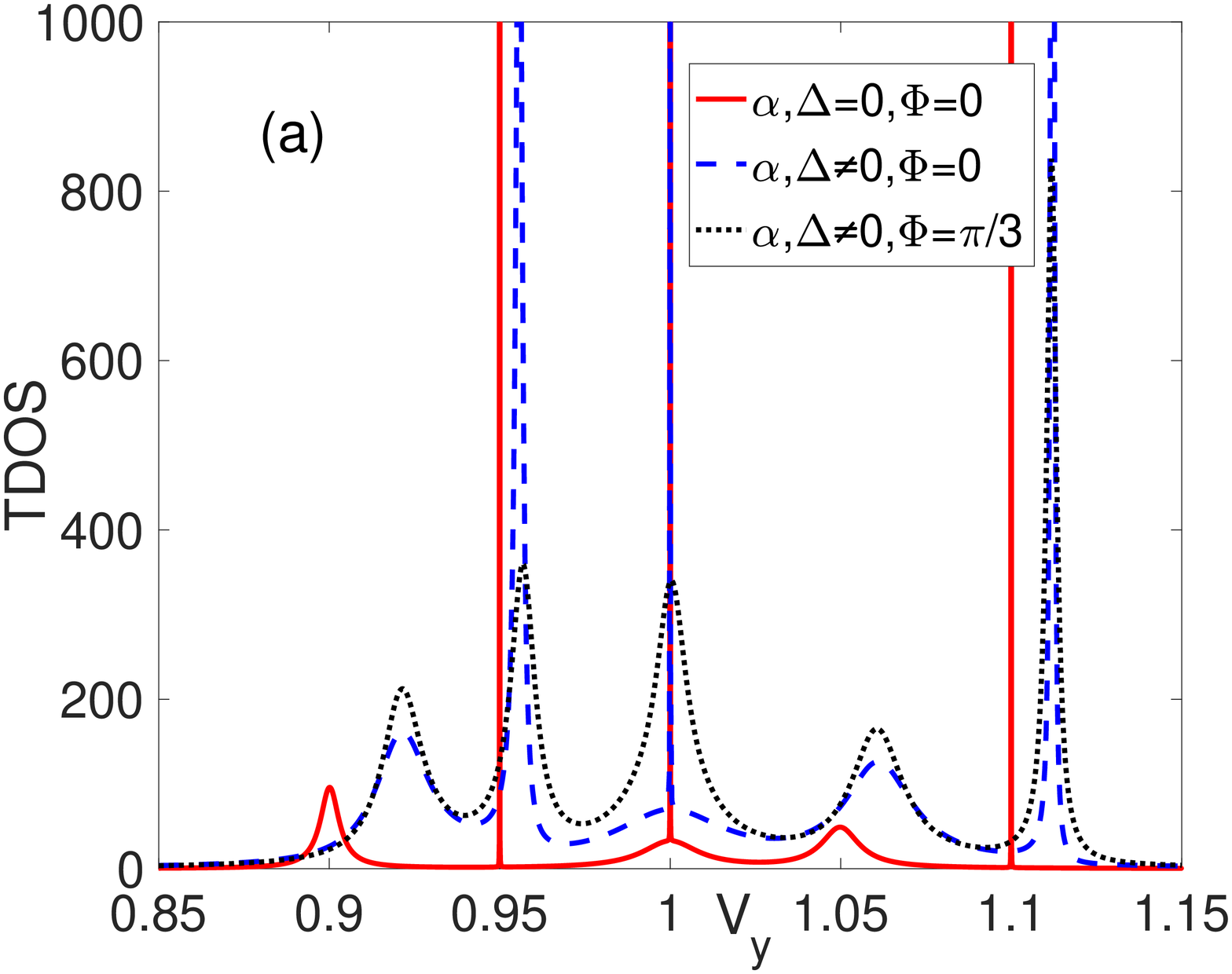}
	\includegraphics[width=0.45\textwidth]{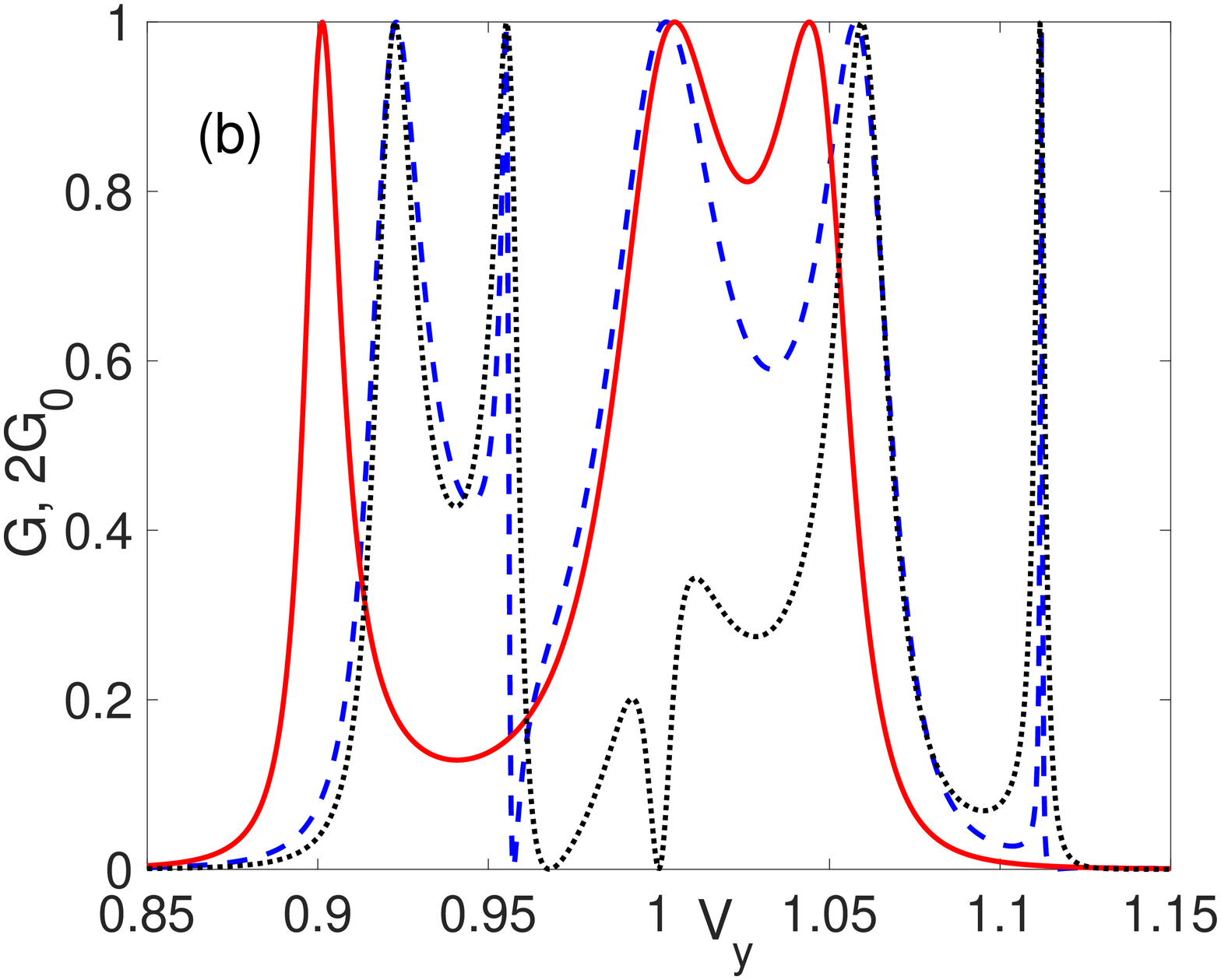}
	\caption{\label{3} (a) The magnetic-field dependence of the TDOS (a) and conductance (b) of the sextuple-QD structure for $t_0=0.1$. The other parameters are the same as in Fig.\ref{2}.}
\end{figure}
As it was shown in the works \cite{volya-03,lu-05,sadreev-06} presence of the degenerate levels ($E_{1,2}$) as well as bonding and antibonding states ($E_{3,4}$ and $E_{5,6}$) for closed system results in the emergence of bound states in continuum (BSCs) if system is open. This concept is illustrated in Fig.\ref{3}a where the total density of states (TDOS), $TDOS\left(\omega=0;~V_{y}\right)=-Tr\left[Im\left\{\hat{G}^{r}\left(\omega=0;~V_{y}\right)\right\}\right]/\pi$, as function of the in-plane magnetic-field energy is plotted. It is evident that for the degenerate states the wide maximum coincides with the BSC (infinitely narrow peak) at $V_y=1$ (see red solid curve). Additionally, two pairs of the separated maximum and BSC corresponding to two pairs of bonding and antibonding states are located on either side of this point. All three wide maxima in the TDOS manifest themselves as the BWRs in the conductance (see red solid curve in Fig.\ref{3}b). In opposite, there are no resonant features associated with the BSCs in the transport properties.

If spin degrees of freedom are taken into account that the combination of SOI and SC pairing gives rise to nonzero coupling between contacts and BSCs with energies $E_{4,6}$. As a result their peaks in the TDOS become broadened and two FRs emerge at the same $V_y$ in the conductance (see blue dashed curves in Fig.\ref{3}a,b). Finally, following the ideas of \cite{lu-05} we can break the system symmetry, e.g. due to the unequal contact-lead couplings or, alternatively, introducing the AB phase. Then all the BSC peaks in the TDOS acquire the finite width and the conductance contains three FRs (see black dotted curves in Fig.\ref{3}a,b). The last result can be realized even in the spinless regime (not shown here). Next, the energy splitting of bonding and antibonding states is defined by $t_0$. In particular, if $t_0=1$ then $t_0\gg t_{1}$ and we get the situation when there are two wide maxima and two BSCs are around $V_y=1$. Additionally, one maximum (BSC) is distantly settled to the left (right) of them.

\begin{figure}[htbp]
	\includegraphics[width=0.5\textwidth]{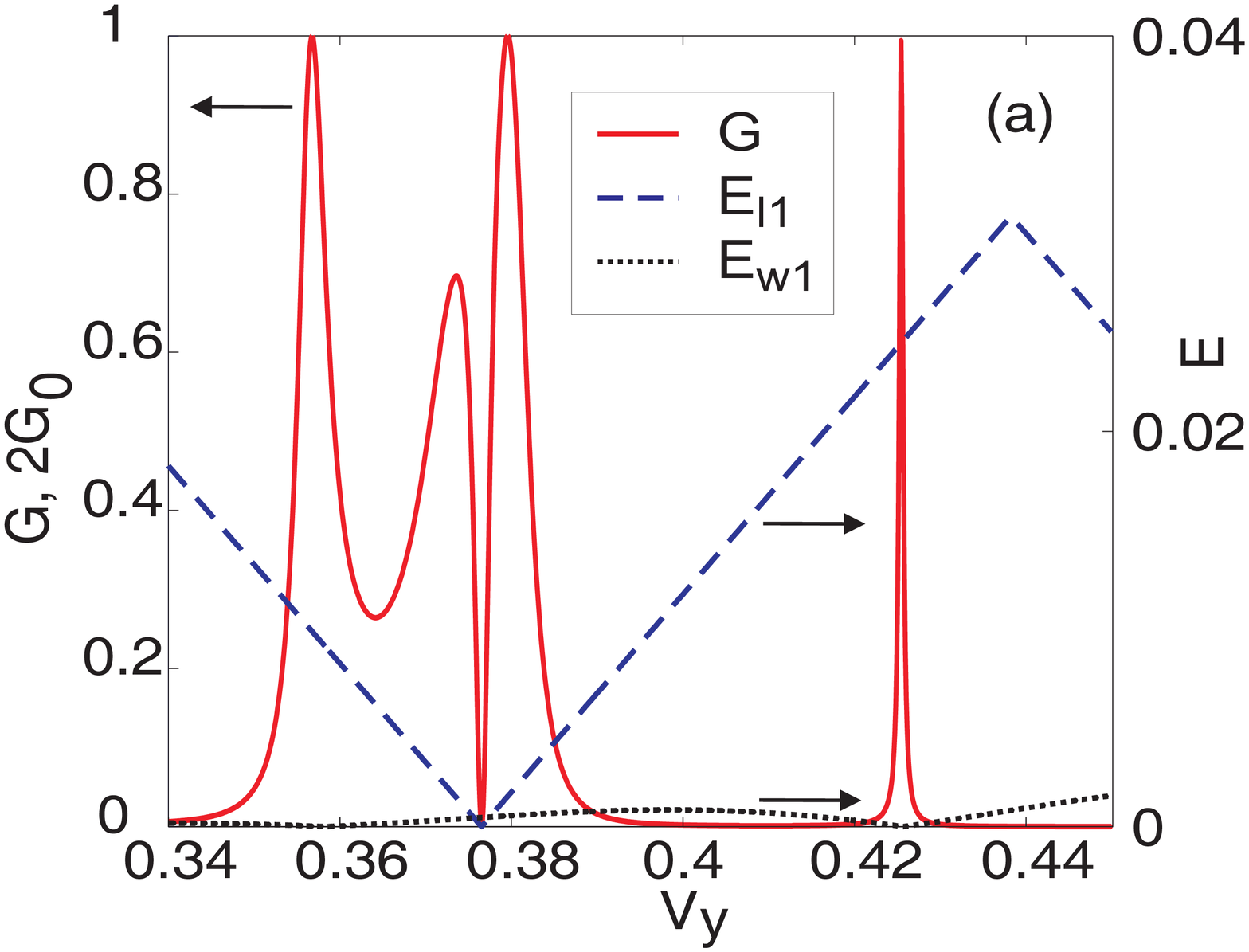}
	\includegraphics[width=0.45\textwidth]{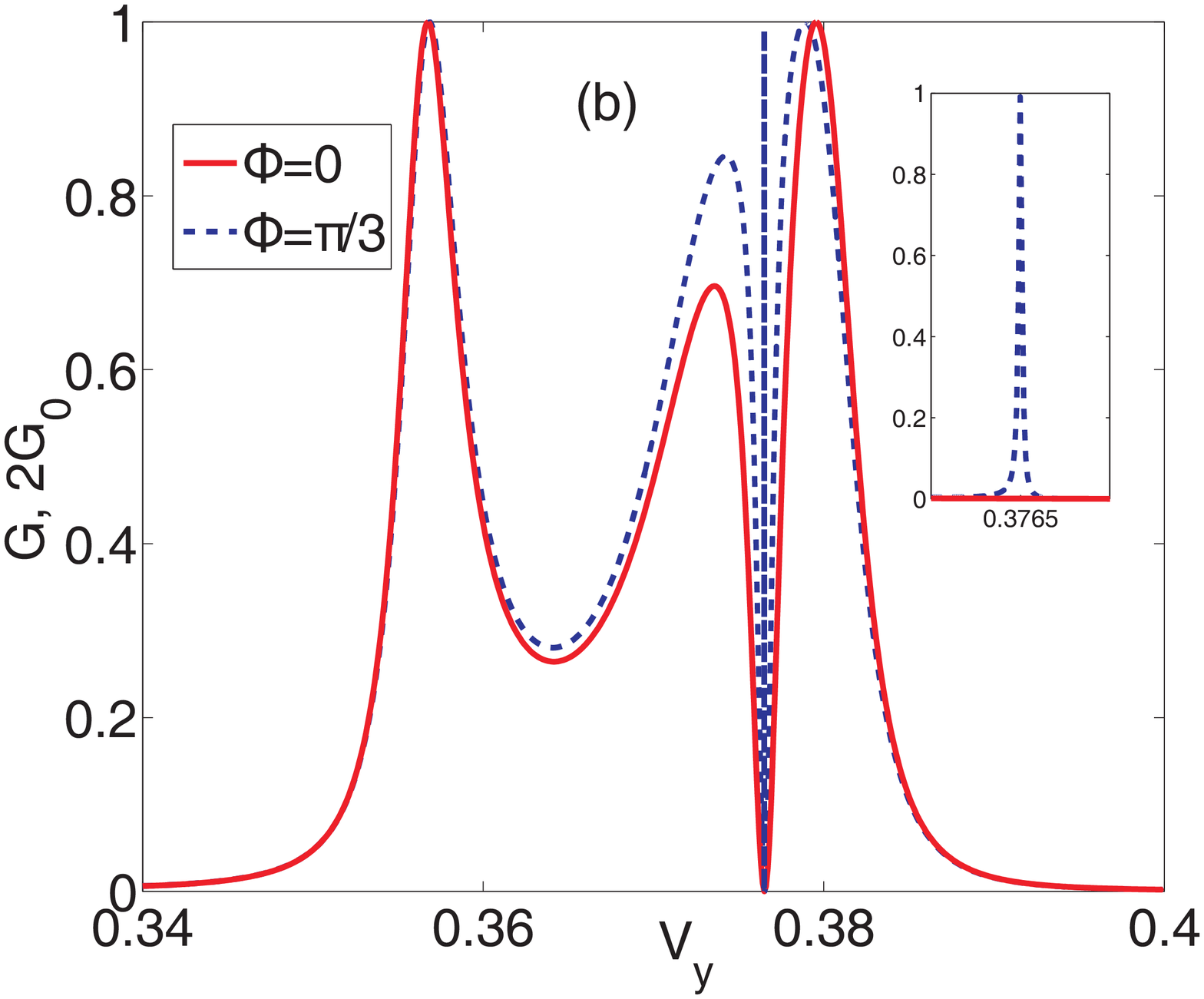}
	\caption{\label{4} (a) The BWRs and FR in the conductance of the AB ring when the SC wire is in the nontrivial phase. The BWR/FR position (the left $y$-axis) coincides with the minimum of the energy $E_{W1}$/$E_{l1}$ (the right $y$-axis). (b) The influence of the AB phase. Inset: the appearance of new FR if $\Phi\neq0$. The parameters are the same as in Fig.\ref{2}.}
\end{figure}
Recurring to the initial system we get the set of such features in the TDOS. However, it is essential that the BSCs obtain finite lifetime only if the wire lowest-energy excitation is close to zero, in other words the topologically nontrivial phase has to be induced (see Fig.\ref{2}a). The typical structure of conductance resonances in this regime is displayed in Fig.\ref{4}a. It is modified in comparison with the above-described picture of the sextuple-QD conductance. In particular, one of the FRs of sextuple-QD structure transforms to the set of BWRs in case of the AB ring. As a result, the FRs close to the $E_{l1}\left(V_y\right)$ minima are only left (one of them is shown in Fig.\ref{4}a). Meanwhile, the BWR positions are substantially defined by the $E_{w1}\left(V_y\right)$ zeros. When $V_{y}$, $\Delta$, $\alpha$ are fixed the BWR- and FR widths depend on $E_{l1}$ and $E_{W1}$, respectively. The higher $E_{l1,W1}$ the narrower the corresponding resonance.

Since the increase of the AB-ring size does not change the system symmetry the degenerate states have to remain. As it was shown above the nonzero AB phase allows to unveil the existence of the corresponding BSC. In Fig.\ref{4}b the new FR with extremely narrow width emerges if $\Phi\neq0$ and it coincides with the $E_{l1}\left(V_y\right)$ zero.

\begin{figure}[htbp]
	\includegraphics[width=0.52\textwidth]{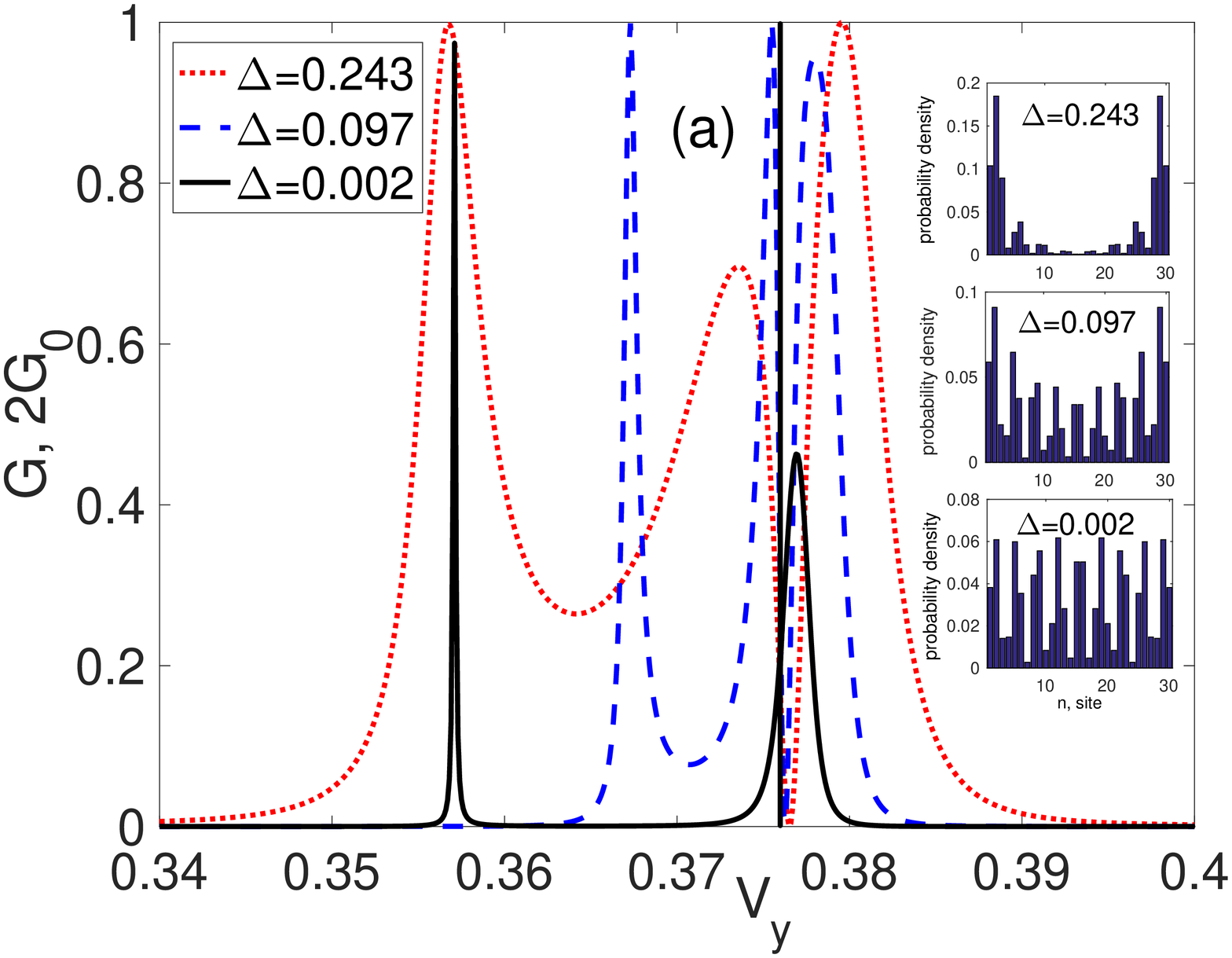}
	\includegraphics[width=0.52\textwidth]{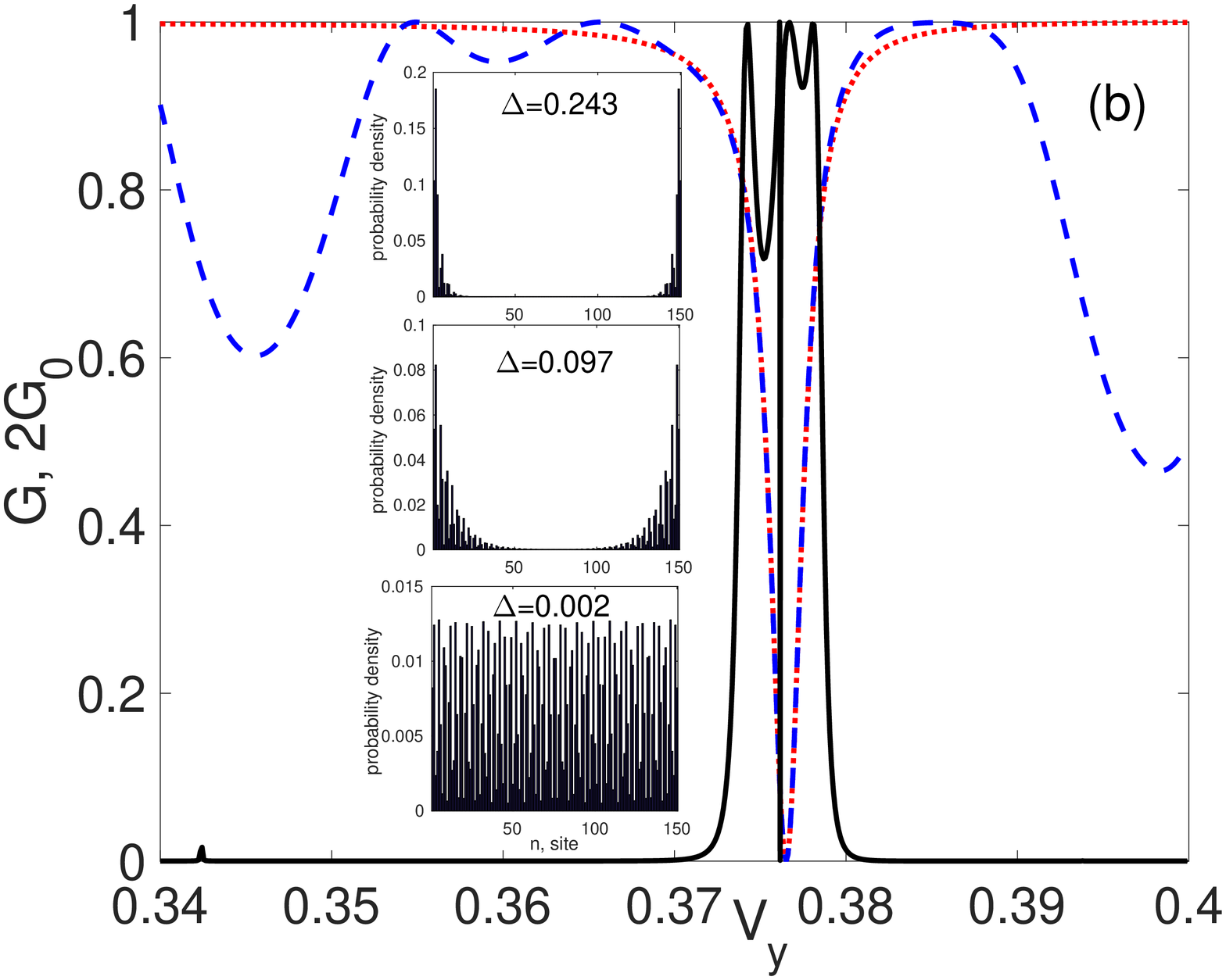}
	\caption{\label{6} The SC-ordering dependence of the FR for $N_W=30$ (a) and $N_W=150$ (b).
		Insets: spatial distributions of the probability density of the wire state with $E_{W1}$ for different $\Delta$ and $V_{y}=0.375$. The other parameters are the same as in Fig.\ref{2}.}
\end{figure}
Thus, the occurrence of various resonances can be qualitatively explained by the fact that different interacting channels can provide the contributions to transport. If it is the wire channel, i.e. $\mu=E_{W1}$, than the BWR appears. If the lead channel, $\mu=E_{l1}$, mainly participates in the transport the Fano-type interference takes place. It is worth to note that similar transport signatures were observed in other nanosize systems containing two interacting channels \cite{lu-05,myoung-18}. 

\subsection{\label{secB}FR properties: MBS vs ABS}
\subsubsection{\label{secB1}The FR width and position}

\begin{figure}[htbp]
    \includegraphics[width=0.5\textwidth]{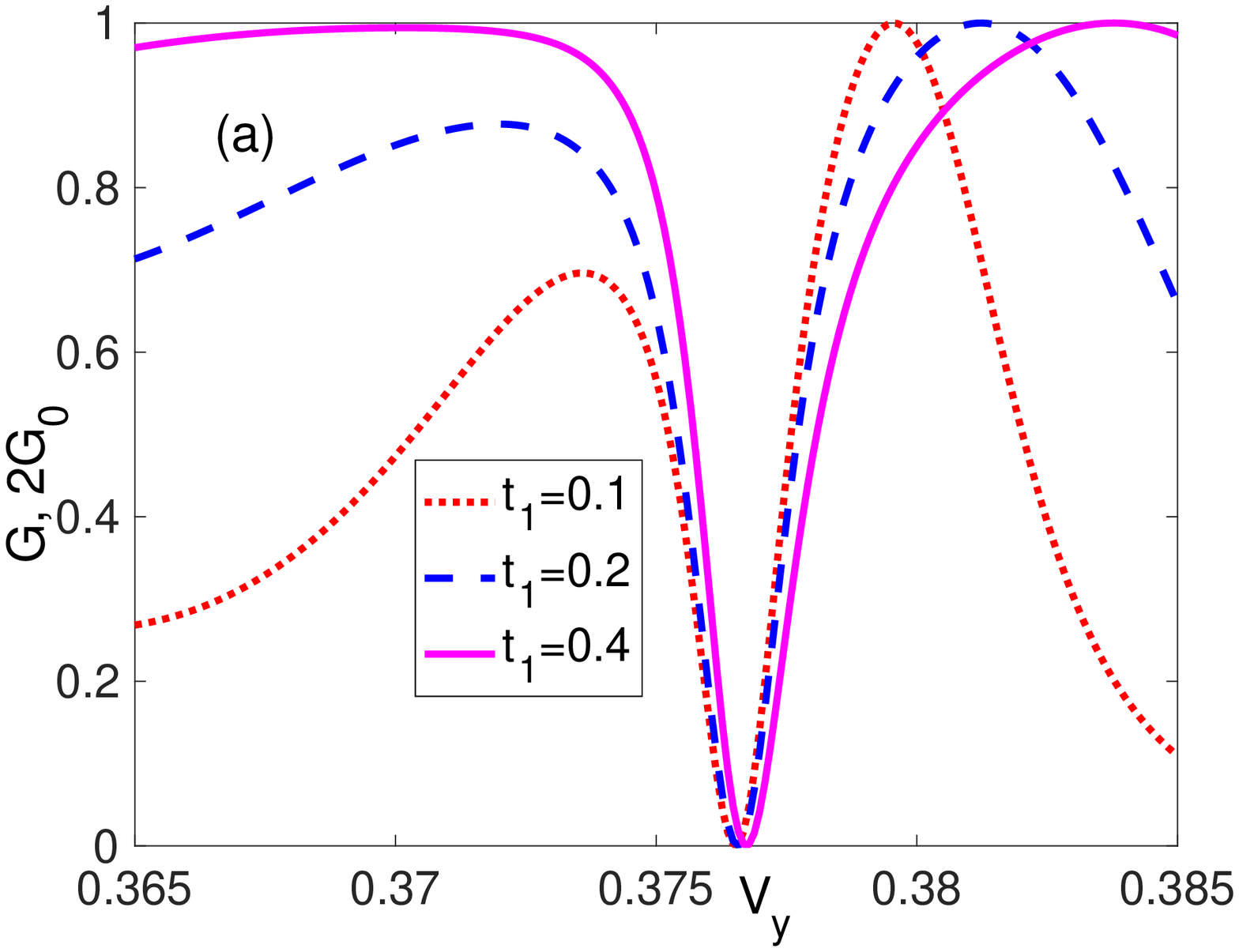}
    \includegraphics[width=0.5\textwidth]{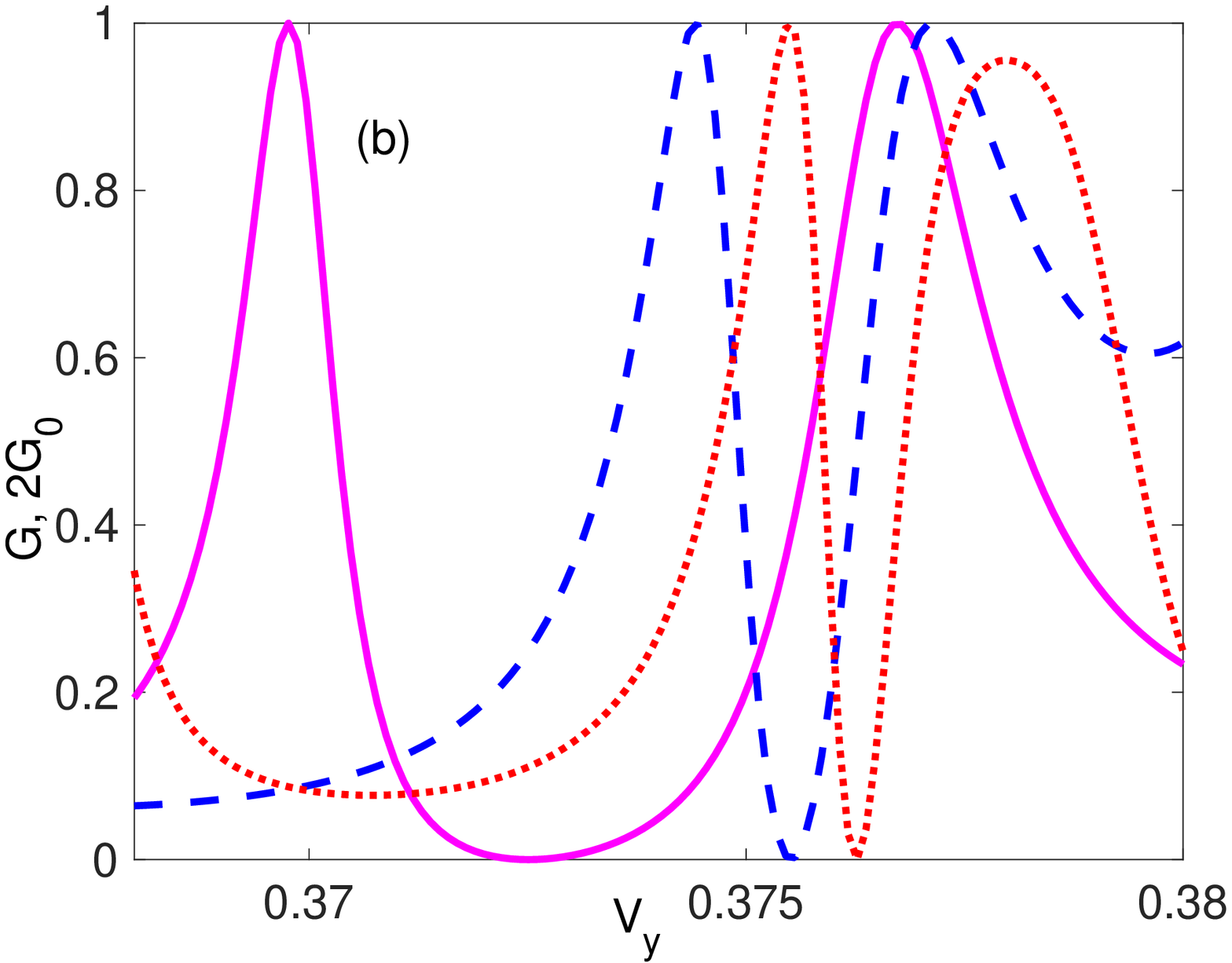}
    \caption{\label{7} The FR-position dependence on the strength of the wire-lead coupling, $t_{1}$, for $\Delta=0.243$ (a) and $\Delta=0.097$ (b). The other parameters are the same as in Fig.\ref{2}.}
\end{figure}
One can note from Figs.\ref{2} and \ref{3} that the FRs appear for any nonzero SOI and SC pairing if the in-plane magnetic field satisfies the inequality $\mu^2+\Delta^2<V_{y}^2<\left(2t - \mu\right)^2+\Delta^2$. The last means that the FR is present irrespective of the type of SC-wire state with $E_{W1}$ since the spatial distribution of the corresponding probability density continuously changes from the edge- (MBS) to bulk (ABS) one as the gap in the SC-wire excitation spectrum decreases \cite{haim-15}. However, the FR properties significantly depend on the type of this state that can be governed by e.g. the in-plane magnetic field or SC-pairing potential. The last case is shown in Fig.\ref{6}. One can compare the FRs at $\Delta=0.243$ and $\Delta=0.097$ (see red dotted and blue dashed curves in Fig.\ref{6}a). The width dramatically decreases as the state with $E_{W1}$ becomes Andreev-type at $\Delta=0.097$ (see the top and middle insets of Fig. \ref{6}a). At the same time, consideration of the longer wire allows to make this state being Majorana-like again at $\Delta=0.097$ that is depicted in the middle inset of Fig.\ref{6}b. As a result, the FR width is not changed as long as the overlap of the MF probability densities gathered at the wire edges is close to zero (see red dotted and blue dashed curves in Fig.\ref{6}b). Finally, the observed FRs collapse (the FR width becomes infinitely small) when $\Delta$ vanishes even though $E_{W1}\approx0$ (see black solid curves in Fig.\ref{6}a and \ref{6}b).

Using the toy model of sextuple-QD structure we demonstrated that the FR positions are defined, in particular, by the hopping parameters. It is interesting that after the increase of number of sites in the system the impact of $t_{1}$ change is greatly determined by the spatial distribution of the SC-wire low-energy state. Despite the growth of the FR width, its position is 'frozen' in case of MBS as $t_{1}$ grows (see Fig.\ref{7}a). Otherwise, when the state is substantially delocalized and transforms to the ABS the increase of $t_{1}$ gives rise to the FR displacement (see Fig.\ref{7}b).

\begin{figure}[htbp]
	   \includegraphics[width=0.5\textwidth]{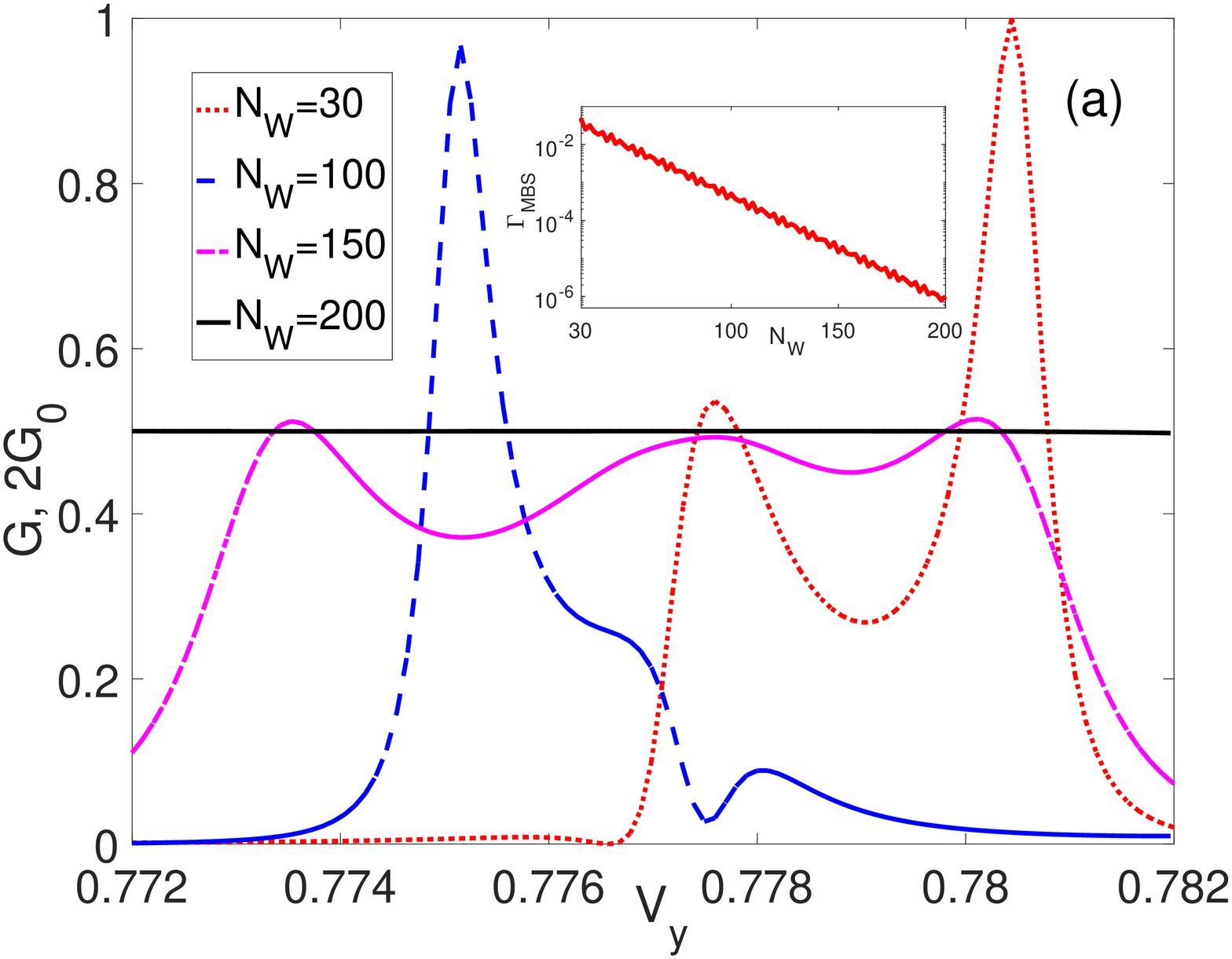}
	   \includegraphics[width=0.5\textwidth]{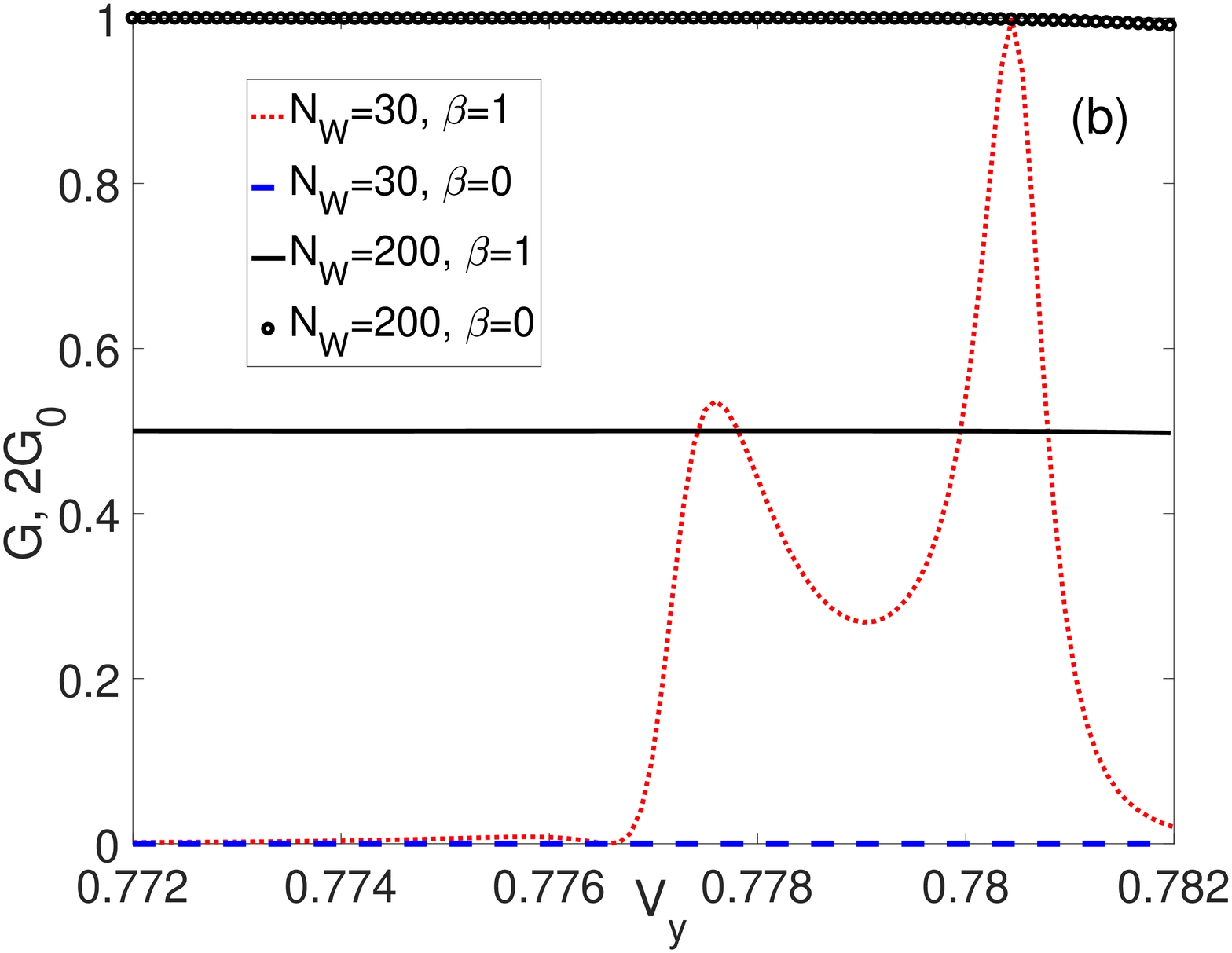}
	\caption{\label{8} The magnetic-field dependence of the conductance in the T-shape transport geometry. (a) The effect of the spatial distribution of the SC-wire low-energy state on the FR. (b) The conductance in the nonlocal ($\beta=1$) and local ($\beta=0$) transport regimes. Inset of Fig.\ref{8}a: the MBS overlap \eqref{olp} as a function of the wire length at $V_y=0.777$. The other parameters are the same as in Fig.\ref{2}.}
\end{figure}

\subsubsection{\label{secB2}T-shape geometry}

The MBS and ABS can be also distinguished due to the modification of the transport scheme. Let us consider the T-shape geometry, i.e. $\Gamma_{LN}=\Gamma_{RN}=0$. In this situation the FR is extremely sensitive to the MBS overlap,
\begin{equation}\label{olp}
\Gamma_{MBS}=\sum\limits_{\sigma;j=N_{W}/2}^{N_{W}/2+1}\left(\left| u_{1j\sigma}\right|^2 +\left| v_{1j\sigma}\right|^2\right),
\end{equation} 
where $u_{1j\sigma},~v_{1j\sigma}$ are the Bogoliubov-transformation coefficients corresponding to the energy $E_{W1}$, site $j$ and spin $\sigma$. This parameter is plotted in the inset of Fig.\ref{8}a. The MBS overlap is reduced by two orders of magnitude if the wire length increases from $N_{W}=30$ to $N_{W}=100$. Simultaneously, the SC-wire low-energy state is modified from the ABS to MBS and the FR minimum placed around $V_{y}=0.777$ rises, i.e. $G\neq0$ (compare red dotted and blue dashed curves in Fig.\ref{8}a). Further elongation of the wire results in the two more orders decrease of $\Gamma_{MBS}$. As a result, the $G_0$-plateau emerges in the conductance (see magenta dash-dotted and black solid curves). 

The reasons of the obtained behavior become clear if we note that the left and right bottom leads in such a T-shaped scheme play a role of side-coupled structures which effective interaction with the contacts is defined by the type of the state with $E_{W1}$. If it is the MBS than these leads are virtually separated from the contacts and the FR does not appear. However, if the state becomes Andreev-type or, in other words, bulk-like than the system is equivalent to the Fano-Anderson model, which describes the interaction between a discrete linear chain and single impurity \cite{kivshar-10}, and the corresponding resonance emerges in the conductance.

Finally, the behavior of local transport via the MBS and ABS is completely different. To demonstrate it we introduce an asymmetry coefficient, $\beta$, such that $t_{RT}=\beta t_{1}$. It is remarkable that in the MBS case the height of the conductance plateau doubles if the transport regime evolves from nonlocal, $\beta=1$, to local, $\beta=0$ (see black solid and circled curves in Fig.\ref{8}b). In turn, the ABS-mediated conductance vanishes in the local transport regime (compare red dotted and blue dashed curves in Fig.\ref{8}b).

The observed features of transport via the MBS and ABS in the T-shape device can be described analytically. In order to achieve this it is essential to notice that transport into the MBS implies tunnel coupling with only one MF. Meanwhile, in the Majorana representation tunneling into the ABS, $\alpha^{+}\left| 0\right\rangle$, brings into play both MFs since $\alpha^{+}=\left(\eta_{1}+i\eta_{2}\right)/2$, where $\eta_{1,2}$ are the MF operators. 

Let us consider these cases separately and start with an effective low-energy Hamiltonian $\hat{H}_{W}=i\xi_0\eta_{1}\eta_{2}/2$ describing the interaction between MFs with intensity $\xi_{0}$. For the sake of simplicity we substitute the four multi-site leads for single-level QDs and consider spinless case, then 
\begin{eqnarray}\label{H_MBS}
&&\hat{H}_{Wl}=t_{1}\left[d_{1}^{+}-d_{1}+\beta\left(d_{2}^{+}-d_{2}\right)\right]\eta_{1}+\nonumber\\
&&~~~~~~~~~~~~~~~~~~~~~~~~~~~~~~~~~~t_{1}\left[d_{3}^{+}-d_{3}+d_{4}^{+}-d_{4}\right]\eta_{2},\nonumber\\
&&\hat{H}_{Cl}=\sum\limits_{k}\left(t_{L}c_{Lk}^{+}d_{1}+t_{R}c_{Rk}^{+}d_{2}+h.c.\right),\\
&&\hat{H}_{l}=\sum\limits_{j=1}^{4}\xi_{j}d_{j}^{+}d_{j},\hat{H}_{i} =\sum\limits_{k}\left(\xi_{k}\mp\frac{eV}{2}\right)c^+_{ik}c_{ik},
~i=L,R.\nonumber 
\end{eqnarray}

Applying \eqref{IL1} for the symmetric system ($t_{L}=t_{R}$) leads to the following expression for the left-contact current:
\begin{equation}\label{ILTsh}
I_{L}=\frac{e\Gamma^2}{4\pi}\int\limits_{-\infty}^{+\infty}d\omega\left(n_{R}-n_{L}\right)\left[2\left| F_{11e}^{r} \right|^2+\left| G_{12e}^{r} \right|^2+\left| G_{12h}^{r} \right|^2 \right],
\end{equation}
where $n_{L\left(R\right)}=n\left(\omega\pm eV/2\right)$; $F^{r}_{11e}\left(\omega\right)$, $G^{r}_{12e}\left(\omega\right)$, $G^{r}_{12h}\left(\omega\right)$ are the Fourier transforms of the anomalous and normal retarded Green's functions,
\begin{eqnarray}
&&F^{r}_{11e}\left(t-t'\right)=-i\Theta\left(t-t'\right)\left\langle d^{+}_{1}\left(t\right) d^{+}_{1}\left(t'\right)\right\rangle, \nonumber\\ &&G^{r}_{12e}\left(t-t'\right)=-i\Theta\left(t-t'\right)\left\langle d_{1}\left(t\right) d^{+}_{2}\left(t'\right)\right\rangle, \nonumber\\ &&G^{r}_{12h}\left(t-t'\right)=-i\Theta\left(t-t'\right)\left\langle d^{+}_{1}\left(t\right) d_{2}\left(t'\right)\right\rangle. \nonumber
\end{eqnarray}
 The term proportional to $F^{r}_{11e}$ in \eqref{ILTsh} describes the contribution from the local Andreev reflection. The last two correspond to the processes of direct charge transfer. Note that the contribution from the crossed Andreev reflection is absent since $\mu_{L\left(R\right)}=\mu\pm eV/2$ \cite{wu-14}.
The linear-response conductance from  \eqref{ILTsh} is equal to
\begin{equation}\label{GLTsh}
G_{L}=G_{0}\frac{\Gamma^2}{2}\left[2\left| F_{11e}^{r}\left(0\right) \right|^2+\left| G_{12e}^{r}\left(0\right) \right|^2+\left| G_{12h}^{r}\left(0\right) \right|^2  \right].
\end{equation}

Utilizing the equation-of-motion technique the sought Green's functions are 
\begin{eqnarray}\label{FG1_MBS}
&&F^{r}_{11e}=-\frac{2t_{1}^2zZ_{B}C_{2}}{zZ_{T}Z_{B}-\xi_{0}^2C_{T}C_{B}}, \\ &&G^{r}_{12e\left(h\right)}=\frac{2\beta t_{1}^2zZ_{B}C_{1h\left(e\right)}C_{2h\left(e\right)}}{z Z_{T}Z_{B}-\xi_{0}^2C_{T}C_{B}},\nonumber
\end{eqnarray}
where $z=\omega+i\delta$, $C_{je\left(h\right)}=z\mp\xi_{j}+i\delta_{jj_T}\Gamma/2$ $\left(j_T=1,2\right)$, $C_{j}=C_{je}C_{jh}$, $C_{T\left(B\right)}=C_{1\left(3\right)}C_{2\left(4\right)}$, $Z_{B}=C_{B}-4t_{1}^2\left(C_{3}+C_{4}\right)$, $Z_{T}=z C_{T}-2t_{1}^{2}\left(C_{1e}+C_{1h}\right)\left(\beta^2C_{1}+C_{2}\right)$.
Then, if $\xi_{1}=\xi_{2}=\xi_{3}=\xi_{4}=0$ than
\begin{equation}\label{FG2_MBS}
G^{r}_{12e}=G^{r}_{12h}=-\beta F^{r}_{11e}=\frac{2\beta t_{1}^{2}\left(z^2-8t_{1}^2\right)}{C_{1e}Z_{1}},
\end{equation}
where $Z_{1}=\left(z^2-8t_{1}^2\right)\left(zC_{1e}-4t_{1}^{2}\left(1+\beta^2\right)\right)-\xi_{0}^2zC_{1e}$.
Thus, for $\beta=1$  the linear-response conductance becomes $G_{L}=G_{0}/2$ and is in agreement with \cite{liu-11}. However, if $\beta=0$ than $G_{L}=G_{0}$. The same results occur if $\xi_{1}=\xi_{2}=\xi_{3}=\xi_{4}\neq0$ and $\xi_{0}=0$ (see Fig.\ref{8}). The conductance \eqref{GLTsh} tends to zero in general case when $\xi_{j},~\xi_{0}\neq0$. 

Next, consider the second case, i.e. the transport via the isolated ABS. Hence, the low-energy wire Hamiltonian is $\hat{H}_{W}=\xi\alpha^{+}\alpha$. As the ABS- and initial second-quantization operators are related by means of the Bogoliubov transformation, $a_{1\left(N_W\right)}\approx u\alpha\pm v\alpha^{+}$, the QD-ABS tunnel Hamiltonian has a following form \cite{tripathi-16}
\begin{eqnarray} \label{HWl_ABS}
&&\hat{H}_{Wl} =t_{1e}\left(d_{1}^{+}+\beta d_{2}^{+}+d_{3}^{+}+d_{4}^{+}\right)\alpha+\\
&&~~~~~~~~~~~~~~~~~~~~+t_{1h}\left(d_{1}+\beta d_{2}+d_{3}+d_{4}\right)\alpha+h.c.,\nonumber
\end{eqnarray}
where $t_{1e\left(h\right)}$ is an electron (hole) tunneling amplitude. 

The solution of equations of motion for the Green's functions for $\xi_{1}=\xi_{2}=\xi_{3}=\xi_{4}=0$ gives
\begin{eqnarray}\label{FG1_ABS}
&&F^{r}_{11e}=-\frac{2t_{1e}t_{1h}z^3}{Z_{2}}, \\ &&G^{r}_{12e\left(h\right)}=\frac{\beta z}{C_{1e}Z_{2}}\left[zC_{1e}\left(t_{1e}^2C_{h\left(e\right)}+t_{1h}^2C_{e\left(h\right)}\right)-\right.\nonumber\\
&&\left.~~~~~~~~~~~~~~~~~~~~~~~-\left(t_{1e}^2-t_{1h}^2\right)^2\left(z\left(1+\beta^2\right)+2C_{1e}\right)\right],\nonumber
\end{eqnarray}
where $C_{e\left(h\right)}=z\mp\xi$,
\begin{eqnarray}\label{Z2}
&&Z_{2}=z^2C_{1e}^2C_{e}C_{h}+\left(t_{1e}^2-t_{1h}^2\right)^2\left(z\left(1+\beta^2\right)+2C_{1e}\right)^2-\nonumber\\
&&~~~~~~~~~~~~~~~~~~-2z^2C_{1e}\left(t_{1e}^2+t_{1h}^2\right)\left(z\left(1+\beta^2\right)+2C_{1e}\right).\nonumber
\end{eqnarray}
It is clearly seen from \eqref{FG1_ABS} that the linear-response conductance vanishes both in nonlocal and local transport via the ABS.
Note that the general formulae for the Green's functions are rather cumbersome to present them explicitly. However, it is necessary to emphasize that in this situation $F^{r}_{11e}\left(0\right)=0$ for $\beta=0,1$ whereas $G^{r}_{12e\left(h\right)}\left(0\right)\neq0$ for $\beta=1$. Such a behavior qualitatively coincides with the numerics in Figs.\ref{8}.  

Thus, the proposed AB ring allows to detect the topological phase transition due to the appearance of FRs. Additionally, analyzing the FR properties one can separate the edge low-energy excitations of the SC wire from the ones having the bulk-like spatial distribution. The last option means that the AB ring is a testbed to distinguish between Majorana and Andreev states.

\subsection{\label{secC}The effects of Coulomb interactions and disorder}

\begin{figure}[htbp]
    \includegraphics[width=0.5\textwidth]{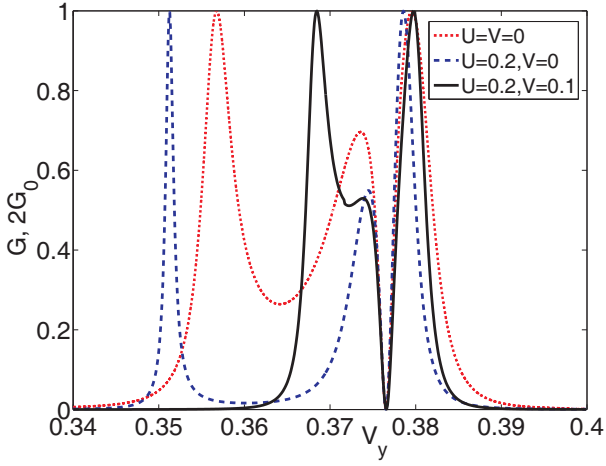}
    \caption{\label{9} The effect of the Coulomb interactions on the BWR and FR. Parameters are the same as in Fig.\ref{2}.}
\end{figure}
To describe the effect of the Coulomb correlations on the obtained features in the conductance we use the generalized mean-field approximation which is valid for the relatively weak interactions, i.e $U,~V \ll t$. It implies the following simplification of the four-operator terms in the wire Hamiltonian \cite{valkov-17}:
\begin{eqnarray}\label{GMFA}
&&Un_{j\uparrow}n_{j\downarrow} = U\left[\langle n_{j\uparrow}\rangle n_{j\downarrow}+\langle n_{j\downarrow}\rangle n_{j\uparrow}-\langle a^{+}_{j\downarrow} a_{j\uparrow}\rangle a^{+}_{j\uparrow} a_{j\downarrow}-\right.\nonumber\\
&&
\left.-\langle a^{+}_{j\uparrow} a_{j\downarrow}\rangle a^{+}_{j\downarrow} a_{j\uparrow}-\langle a^{+}_{j\uparrow} a^{+}_{j\downarrow}\rangle a_{j\uparrow} a_{j\downarrow}-\langle a_{j\uparrow} a_{j\downarrow}\rangle a^{+}_{j\uparrow} a^{+}_{j\downarrow}\right],\nonumber\\
&&V\sum\limits_{\sigma\sigma'}n_{j\sigma}n_{j+1,\sigma'}=V\sum\limits_{\sigma\sigma'}
\left[\langle n_{j\sigma}\rangle n_{j+1,\sigma'}+\langle n_{j+1,\sigma}\rangle n_{j\sigma}-\right.\nonumber\\
&&\left.-\langle a^{+}_{j+1,\sigma} a_{j\sigma'}\rangle a^{+}_{j\sigma'} a_{j+1,\sigma}-\langle a^{+}_{j\sigma} a_{j+1,\sigma'}\rangle a^{+}_{j+1,\sigma'} a_{j\sigma}-\right.\nonumber\\
&&\left.-\langle a^{+}_{j\sigma} a^{+}_{j+1,\sigma'}\rangle a_{j\sigma} a_{j+1,\sigma'}-\langle a_{j\sigma} a_{j+1,\sigma'}\rangle a^{+}_{j\sigma} a^{+}_{j+1,\sigma'}\right].
\end{eqnarray}
The normal and anomalous averages in \eqref{GMFA} can be calculated self-consistently employing the spin-dependent coefficients of the Bogoliubov transformation, $u_{li\sigma},~v_{lj\sigma}$,
\begin{eqnarray}\label{avs}
&&\langle a^{+}_{i\sigma} a_{j\sigma'}\rangle = \sum\limits_{l=1}^{2N_W}\left[f_{l}u_{li\sigma}u^{*}_{lj\sigma'}+
\left(1-f_{l}\right)v^{*}_{li\sigma}v_{lj\sigma'}\right],\nonumber\\
&&\langle a_{i\sigma} a_{j\sigma'}\rangle = \sum\limits_{l=1}^{2N_W}\left[f_{l}v_{li\sigma}u^{*}_{lj\sigma'}+
\left(1-f_{l}\right)u^{*}_{li\sigma}v_{lj\sigma'}\right],
\end{eqnarray}
where $f_{l}$ - the Fermi distribution function of the quasiparticle state with the energy $E_{l}$. As a consequence of the expansion \eqref{GMFA}, the matrix of the wire Hamiltonian, $\hat{H}_{W}$, contains both the corrections of the nonzero elements and new terms. In particular, the last ones are responsible for the on-site spin-flip processes and intersite SC pairing.

The influence of the Coulomb interactions on the FR is displayed in Fig.\ref{9}. It is seen that the on-site correlations slightly shift the maximum of the FR but the antiresonance remains at the same place (compare red dotted and blue dashed curves). Simultaneously, the BWR indicating the pure tunneling via the MBS moves to the left because of the corresponding shift of the MBS zero energy when $U\neq0$. If now $V\neq0$ the BWR goes to the opposite direction (see black solid curve). But the FR is stuck at the same place and its characteristics are changed negligible. As it follows from the used parameters the discussed results are obtained in the regime of low carrier concentration in the wire. At the high carrier-concentration regime ($\mu > t+\sigma V_{y}+U\langle a^{+}_{j\bar{\sigma}} a_{j\bar{\sigma}}\rangle$) the observed behavior is qualitatively similar.

\begin{figure}[htbp]
	\includegraphics[width=0.5\textwidth]{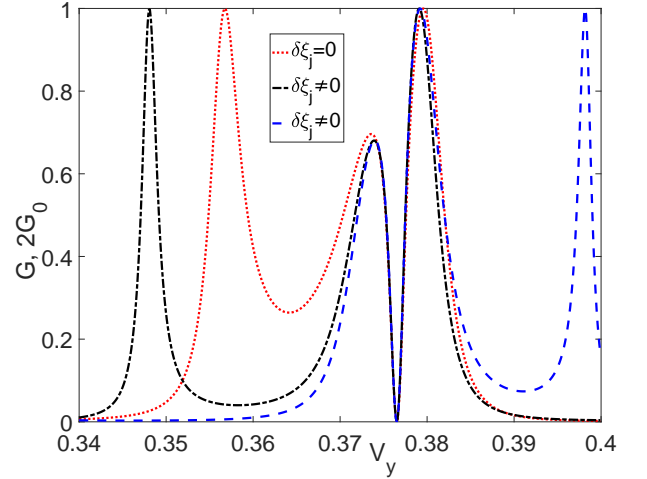}
	\caption{\label{10} The effect of diagonal disorder on the BWR and FR. Parameters are the same as in Fig.\ref{2}.}
\end{figure}
The effect on the linear-response conductance of the AB ring similar to the Coulomb interactions occurs if the diagonal disorder  is present in the SC wire, $\xi_{\sigma}+\delta\xi_{j},~j=1,...,N_{W}$ (where $\delta\xi_{j}$ is an jth site random potential which takes values with an uniform distribution in the interval $\left[-1/2,~1/2\right]$). In Fig.\ref{10} the influence of this issue is shown for two independent distributions of $\delta\xi_{j}$ (see blue dashed and black dash-dotted curves).  

\section{\label{sec5}Conclusion}

In the article we investigated the linear-response conductance of the AB ring focusing on the interplay between the internal energy structures of the AB arms (leads) and SC wire which bridges the arms. Employing the nonequilibrium Green's functions we showed that the BWRs and FRs emerge when the SC wire is switched in the topologically nontrivial phase by the in-plane magnetic field. Using the sextuple-QD toy model we demonstrated that the FRs are related to the dependence of BSCs' lifetime on hopping parameters as well as the SOI and SC pairing in the wire. Connection between the type of the lowest-energy state in the SC wire, the MBS or ABS, and the FR properties is exhibited. Additionally, it is shown that the spatial distribution of this state can be examined via the T-shaped transport scheme. Thus, taking into account the ongoing debates on the ways to distinguish between the MBS and ABS the obtained FR can be used as a correspoinding tool (in favor of the MBS). Our results provide additional data for new experiments aiming to observe the MBS manifestation in coherent transport. 

\begin{acknowledgments}
The reported study was funded by the RAS Presidium program for fundamental
research No. 32 "Nanostructures: physics, chemistry, biology and technology fundamentals", Russian Foundation for Basic Research (projects Nos. 16-02-00073, 17-02-00135, 18-32-00443), Government of Krasnoyarsk Territory, Krasnoyarsk Regional Fund of Science to the research projects: "Majorana bound fermions in the nanomaterials with strong electron correlations and quantum electron transport in the devices containing these materials" (No. 17-42-240441), "The manifestation of Coulomb interactions and effects of bounded geometry in the properties of topological edge states of nanostructures with spin-orbit interaction" (No. 18-42-243017). S.A. is grateful to the Council of the President of the Russian Federation for Support of Young Scientists and Leading Scientific Schools, project No. MK-3722.2018.2. M.Yu.K. thanks the Program of Basic Research of the National Research University Higher School of Economics for support.
\end{acknowledgments}

\bibliography{Majorana}

\end{document}